\documentclass{jpp}
\usepackage{graphicx}
\usepackage{mathtools}
\usepackage[utf8]{inputenc}
\usepackage[T1]{fontenc}
\usepackage{xcolor}
\usepackage{amsmath}
\usepackage{lineno}
\usepackage[hypertexnames=false]{hyperref}

\newcommand{\fley}[1]{\textcolor{blue}{#1}}

\shorttitle{Relativistic double-adiabatic equations}
\shortauthor{F. Ley, A. Tran, and E. G. Zweibel}

\title{On the double-adiabatic equations in the relativistic regime}

\author{Francisco Ley\aff{1}
  \corresp{\email{fley.astro@gmail.com}}, Aaron Tran\aff{3}, Ellen G. Zweibel\aff{2,3}
 }

\affiliation{
\aff{1}Departamento de Física, Facultad de Ciencias Físicas y Matemáticas, Universidad de Chile, Santiago, RM 8370449, Chile

\aff{2}Department of Astronomy, University of Wisconsin–Madison, 475 N. Charter Street, Madison, WI 53706, USA

\aff{3}Department of Physics, University of Wisconsin–Madison, 1150 University Avenue, Madison, WI 53706, USA
}

\begin{document}
 
\maketitle

\begin{abstract}
We revisit the double adiabatic evolution equations and extend them to the relativistic and ultrarelativistic regimes. We analytically solve the relativistic, time-dependent drift kinetic equation for a homogeneous, magnetized, collisionless plasma and obtain a solution explicitly dependent on the magnetic field and density variations. In the case of an initial relativistic Maxwellian distribution, a natural extension to an anisotropic Maxwell-Jüttner is obtained. We calculate the moments of this time-dependent solution and obtain analytical expressions for the evolution of the perpendicular and parallel pressures in the ultrarelativistic case. We numerically solve the moment equations in the relativistic case and obtain general expressions for the double-adiabatic equations in this regime. We confirm our results using fully kinetic particle-in-cell simulations of shearing and compressing boxes. Our findings can be readily applied to relativistic species including cosmic-rays and electron-positron pairs, present in astrophysical plasmas like pulsar wind nebulae, astrophysical jets, black hole accretion flows, and Van Allen radiation belts.
\end{abstract}


\section{Introduction}\label{sec:introduction}

A diversity of astrophysical phenomena host hot, diluted, and turbulent plasmas where ambient conditions make Coulomb collisions between particle species scarce, therefore being weakly collisional or collisionless. This absence of collisions make these systems prone to departures from thermodynamic equilibrium, giving rise to a plethora of kinetic scale phenomena that mediates the momentum and energy transport between scales. In this sense, kinetic scale processes become key to understand the global thermodynamic evolution of these astrophysical settings. Some examples of astrophysical scenarios where these plasmas are present are the solar wind, low luminosity accretion disks around supermassive black holes (SMBH, e.g. Sgr A$^*$ at the center of our Galaxy, and the SMBH at the center of M87), and the intracluster medium (ICM) of galaxy clusters. In addition to these scenarios, there are systems where these collisionless plasmas become relativistically hot. Some examples of this type include pulsar wind nebulae, relativistic jets from active galactic nuclei (e.g. blazars, radio-loud quasars), and cosmic rays. 

An important feature of magnetized, collisionless systems is the development of a pressure anisotropy $\Delta P_{\alpha} \equiv P_{\perp,\alpha}-P_{\parallel,\alpha}$, where $P_{\perp,\alpha}$, $P_{\parallel,\alpha}$ are the components of the pressure tensor of species $\alpha$ perpendicular and parallel to the ambient magnetic field. This anisotropy in the pressure arises from the conservation of the particle adiabatic invariants $\mathcal{M} = p_{\perp,\alpha}^2/2B$, so called magnetic moment, and $\mathcal{J} = \oint p_{\parallel,\alpha}d\ell$, the longitudinal action, where $p_{\perp,\alpha}$, $p_{\parallel,\alpha}$ are the perpendicular and parallel components of the momentum of species $\alpha$ , and the integral of the parallel action is taken between points of concentration of magnetic field lines, or so called bounce points. Written in this form, these invariants are conserved in both nonrelativistic and relativistic regimes. The evolution of $P_{\perp,\alpha}$ and $P_{\parallel,\alpha}$ then decouples, and each component evolve independently. 

The first description for the adiabatic evolution of the perpendicular and parallel pressures in a collisionless plasma was given by the influential paper by  \citet{CGL1956}, now known as double-adiabatic or CGL equations. These relations have been proven useful in a variety of contexts, e.g., for constructing fluid models with anisotropic closures to MHD equations (e.g. \citet{Hunana2019, Majeski2024}), as well as for contributing to understanding the temperature evolution in the solar wind \citep{Matteini2007, Matteini2013}. On the other hand, the validity of the CGL model is delimited by the validity of the conservation of adiabatic invariants. There exists a variety of kinetic processes that can break the adiabatic invariance, e.g., Coulomb collisions, heat fluxes, and pressure anisotropy driven microinstabilities. This way, CGL models have been one important first step into the richness of kinetic plasma processes.

Despite their great success, CGL equations are only valid in the nonrelativistic regime. Relativistic extensions to the double-adiabatic equations were proposed in \citet{Newcomb1982,HolmKuperschmidt1986,Gedalin1991}, and \citet{GedalinOiberman}. In \citet{Gedalin1991} and \citet{GedalinOiberman} , most general state equations for an anisotropic collisionless plasmas were derived. In these equations, $P_{\perp}$ and $P_{\parallel}$ are expressed as a function of the density, magnetic field strength, and the internal energy of the plasma $\varepsilon$, the latter calculated as the appropriate moment of a general, gyrotropic distribution function $f(p_{\perp}^2,p_{\parallel}^2)$, solution of the Vlasov equation for a well magnetized plasma, although no specific functional form for $f(p_{\perp}^2,p_{\parallel}^2)$ was provided.

In this work, we directly solve the drift kinetic equation analytically, and obtain a general solution for a time-dependent, gyrotropic distribution function $f(t,p_{\perp},p_{\parallel})$, for a given initial, well-behaved distribution $f_0(p_{\perp},p_{\parallel})$. We then consider three different cases as initial conditions: nonrelativistic Maxwellian, relativistic and ultrarelativistic Maxwell-Jüttner thermal distributions, and calculate moments to obtain evolution equations for $P_{\perp}$ and $P_{\parallel}$ in each of the three different regimes. We directly recover the CGL equations for the nonrelativistic case, and obtain new double-adiabatic evolution for the relativistic and ultrarelativistic cases, dependent on the density $n(t)$ and field strength $B(t)$ only. We confirm that the obtained double-adiabatic equations satisfy the general state equations proposed in \citet{Gedalin1991} and \citet{GedalinOiberman} in both relativistic and ultrarelativistic regimes. We perform particle-in-cell (PIC) simulations with shearing and compressing drivings to confirm our analytical results, and obtain remarkable agreement in both relativistic and ultrarelativistic regimes. The time-dependent distribution function obtained in the relativistic regime also constitutes a natural extension to an anisotropic Maxwell-Jüttner, derived directly as a solution of the Vlasov equation. We confirmed that this anisotropic Maxwell-Jüttner agrees with PIC simulations notably well.

A similar method of solving the Vlasov equation was presented in \citet{Zhdankin2023} for the case of an ultrarelativistic plasma with synchrotron cooling electrons. Incidentally, \citet{Zhdankin2023} also presented a type of anisotropic distribution for ions, and obtained approximate relations for $P_{\perp}$ and $P_{\parallel}$ for a weak magnetic field strength. In this work we consider any level of anisotropy and magnetic field strength, allowing the results to be completely general. 

This paper is organized as follows. In section \ref{sec:theoreticalBasis} we present the main results of this work: we analytically solve the drift kinetic equation, and obtain time-dependent solutions for nonrelativistic, relativistic, and ultrarelativistic initial distributions. We then calculate moments of these solutions and obtain double-adiabatic evolution equations for $P_{\perp}$ and $P_{\parallel}$ in each regime. In section \ref{sec:simulationsetup} we describe our simulation methods and setup for our shearing and compressing PIC simulations. In section \ref{sec:simulationresults} we compare the analytical results of section \ref{sec:theoreticalBasis} with PIC simulations of both shearing and compressing drivings, for mildly relativistic and ultrarelativistic initial temperatures. We discuss possible future directions that this work can open in section \ref{sec:Discussion}. In section \ref{sec:conclusions} we summarize our results and present our conclusions.

\section{Theoretical Basis}\label{sec:theoreticalBasis}

Consider a periodic, uniform, magnetized, collisionless flux tube of plasma of radius $r$ and length $l$ with no scattering mechanism active and zero heat fluxes. The evolution of the particle distribution function $f_{\alpha}(t,p_{\perp},p_{\parallel})$, which we assume gyrotropic, is given by the drift kinetic equation \citep{Kulsrud1983, Kulsrud2005}

\begin{align}
    \frac{\partial f_{\alpha}}{\partial t} + \frac{dp_{\perp}}{dt}\frac{\partial f_{\alpha}}{\partial p_{\perp}} + \frac{dp_{\parallel}}{dt}\frac{\partial f_{\alpha}}{\partial p_{\parallel}} = 0,
    \label{eq:DKE_raw}
\end{align}

where $p_{\perp}$,$p_{\parallel}$ are the perpendicular and parallel momenta with respect to the direction of the mean magnetic field, and $\alpha$ denotes the particle species. We will assume that the density $n_{\alpha}$ and magnetic field strength $B$ vary very slowly in both spatial and temporal scales: $L_n s_n, L_B s_B \gg v_{th,\alpha}$, where $L_n \equiv n/\nabla n$, $L_B \equiv B/\nabla B$, $s_n \equiv d\ln n_{\alpha}/dt$, $s_B \equiv d\ln B/dt$, and $v_{th,\alpha}$ is the thermal velocity of the species $\alpha$. This way, particle free streaming cannot transport away variations of $n$ and $B$, and we can neglect the spatial derivatives of $f$. The long parallel transit times implied under this assumption allow us to also neglect heat fluxes. We note that a complete, self consistent treatment will have to include the convective derivative of $f_{\alpha}$. Particle drifts that can lead to finite off-diagonal terms in the pressure tensor and heat fluxes can, in general, exists, but consideration of them is beyond the scope of the present work.

Assuming that the spatial scales are much larger than the particle's gyroradius, $n/\nabla n$, $B/\nabla B$ $\gg R_{L,\alpha}$, and the the plasma is well magnetized, $\Omega_{\alpha}/s \gg 1$, where $\Omega_{\alpha} = eB/m_{\alpha}c^2$ is the nonrelativistic cyclotron frequency of species $\alpha$, the following quantities become adiabatic invariants of the system: the particle magnetic moment $\mathcal{M} = p_{\perp}^2/2B $ and the longitudinal action $\mathcal{J} = \oint p_{\parallel} d\ell$, where the integral is taken between periodic points within the flux tube, and it becomes an invariant provided that the variation of the orbit is not substantial during a particle transit through the flux tube\footnote{The adiabaticity associated to $\mathcal{J}$ can actually be traced back to symmetries and conservation of the distribution function through phase space, and is not tight to the particles actually having an adiabatic invariant. In this sense, both trapped and untrapped particles are double-adiabatic, and the argument can be more general than just within a flux tube (see \citet{Wierzchucka2026}).  }. We can then use the conservation of magnetic moment, longitudinal action, particle number and magnetic flux to replace the time derivatives of $p_{\perp},p_{\parallel}$ above, and obtain (see e.g. \citet{Lichko2017})

\begin{align}
    \frac{\partial f_{\alpha}}{\partial t} + \frac{\dot{B}}{2B}p_{\perp}\frac{\partial f_{\alpha}}{\partial p_{\perp}} + \left( \frac{\dot{n}_\alpha}{n_{\alpha}} - \frac{\dot{B}}{B} \right) p_{\parallel}\frac{\partial f_{\alpha}}{\partial p_{\parallel}} = 0.
    \label{eq:DriftKineticEquationCyl}
\end{align}

Equation \eqref{eq:DriftKineticEquationCyl} is valid in both nonrelativistic and relativistic regimes. If we go to spherical coordinates in momentum, $p = (p_{\perp}^2 + p_{\parallel}^2)^{1/2} $ and $ \mu = p_{\parallel}/p $, the drift kinetic equation reads:

\begin{align}
    \frac{\partial f_{\alpha}}{\partial t} + \left( \frac{\dot{n}_{\alpha}}{3n_{\alpha}}p + \left[ \frac{2}{3} \frac{\dot{n}_{\alpha}}{n_{\alpha}} - \frac{\dot{B}}{B}\right]\frac{1}{2}(3\mu^2 - 1)p\right)\frac{\partial f_{\alpha}}{\partial p} + \frac{3}{2}\left[ \frac{2}{3} \frac{\dot{n}_{\alpha}}{n_{\alpha}} - \frac{\dot{B}}{B}\right]\mu (1 - \mu^2)\frac{\partial f_{\alpha}}{\partial \mu} = 0. 
    \label{eq:DriftKineticEquationSph}
\end{align}

We remind the reader that our choice of variable $\mu$ corresponds to the cosine of the pitch-angle of the particle, not to be confused with the magnetic moment $\mathcal{M}$. Note that the term in square brackets can be written as $(2/3)\dot{n}_{\alpha}/n_{\alpha} - \dot{B}/{B} = -\hat{b}\hat{b}:\nabla \textbf{u}_{\alpha} + (1/3)\nabla \cdot \textbf{u}_{\alpha}$, and $(1/3)\dot{n}_{\alpha}/n_{\alpha} = -(1/3)\nabla \cdot \textbf{u}_{\alpha}$, so this combination of $n$ and $B$ variations effectively represents any external motion the plasma is subjected to. For convenience, define $R \equiv n_{\alpha}^{-2/3}B$, then we can simplify equation \eqref{eq:DriftKineticEquationSph} to get

\begin{align}
    \frac{\partial f_{\alpha}}{\partial t} + \left( \frac{\dot{n}_{\alpha}}{3n_{\alpha}}p - \frac{\dot{R}}{R}p\frac{1}{2}(3\mu^2 - 1) \right)\frac{\partial f_{\alpha}}{\partial p} - \frac{3}{2}\frac{\dot{R}}{R}\mu (1 - \mu^2)\frac{\partial f_{\alpha}}{\partial \mu} = 0.
    \label{eq:DriftKineticEquationR}
\end{align}

Given an initial distribution $f_{\alpha,0}(p,\mu)$ at some initial time $t=t_0$, equation \eqref{eq:DriftKineticEquationR} can be promptly solved by the method of characteristics to obtain a solution for $f_{\alpha}(t,p,\mu)$. Recognizing that $f_{\alpha}(t,p,\mu) = f_{\alpha,0}(p_0,\mu_0)$ and 

\begin{align}
    \nonumber
    p_0 &= p\left(\frac{n_{\alpha}}{n_{\alpha,0}}\right)^{-1/3}\left\{ \left(\frac{R}{R_0} \right)^{-1} + \mu^2\left[ \left(\frac{R}{R_0}\right)^2 - \left(\frac{R}{R_0}\right)^{-1}\right] \right\}^{1/2}\\
    \nonumber
    \mu_0 &= \left[ 1 + \frac{1-\mu^2}{\mu^2}\left(\frac{R}{R_0}\right)^{-3} \right]^{-1/2},
\end{align}

we obtain

\small{%
\begin{align}
    \nonumber
    &f_{\alpha}(t,p,\mu) = \\ &f_{\alpha,0}\left( p\left(\frac{n_{\alpha}}{n_{\alpha,0}}\right)^{-1/3}\left\{ \left(\frac{R}{R_0} \right)^{-1} + \mu^2\left[ \left(\frac{R}{R_0}\right)^2 - \left(\frac{R}{R_0}\right)^{-1}\right] \right\}^{1/2} \ , \ \left[ 1 + \frac{1-\mu^2}{\mu^2}\left(\frac{R}{R_0}\right)^{-3} \right]^{-1/2}  \right),
    \label{eq:SolutionDistributionFunctionSph}
\end{align}
}

\normalsize
where $R_0 = n_{\alpha,0}^{-2/3}B_0$ are the initial values of $n_{\alpha}(t)$, $B(t)$ at $t=t_0$. Given an initial distribution function $f_{\alpha,0}(p,\mu)$, equation \eqref{eq:SolutionDistributionFunctionSph} then constitutes a completely general, time-dependent, homogeneous distribution function evolving in response to any specific $B(t)$, $n_{\alpha}(t)$ variation. In this work, we will consider three different cases for $f_{\alpha,0}(p,\mu)$: nonrelativistic Maxwell-Boltzmann, relativistic Maxwell-Jüttner, and ultrarelativistic Maxwell-Jüttner distributions.

One important feature of solution \eqref{eq:SolutionDistributionFunctionSph} will be the development of a pressure anisotropy $P_{\perp} \neq P_{\parallel}$, where $P_{\perp}$, $P_{\parallel}$ are the pressures perpendicular and parallel to the direction of the mean magnetic field $\textbf{B}$. We will see that, depending on the choice of $f_{\alpha,0}(p,\mu)$, we can obtain evolution equations for $P_{\perp}(t)$ and $P_{\parallel}(t)$ as a function of $B(t)$ and $n(t)$. These evolution equations will recover the classic CGL double-adiabatic equations \citep{CGL1956} only in the nonrelativistic regime, i.e., when considering a Maxwellian distribution for $f_{\alpha,0}(p,\mu)$, and will depart from CGL when going to the relativistic and ultrarelativistic regimes (e.g. when considering a Maxwell-Jüttner distribution). Relevant additional studies using the
solution distribution function \eqref{eq:SolutionDistributionFunctionSph}, such as a linear stability analysis and different initial $f_{\alpha,0}(p,\mu)$, will be deferred to future studies.

\subsection{Non-relativistic case: Maxwell-Boltzmann distribution}
\label{sec: Nonrelativistic}
Let us benchmark our results by considering first the case when the initial distribution $f_{0}$ is a Maxwellian distribution function

\begin{align}
    f_{0}(v) = \frac{n_0}{\pi^{3/2}v_{th}^3}\exp\left(-\frac{v^2}{v_{th}^2}\right),
\end{align}

where $v_{th} = \sqrt{2k_BT_0/m}$ is the thermal velocity and $T_{0}$ is the temperature, and we passed from momentum to velocity coordinates owing to the nonrelativistic regime. For simplicity, let us drop the index $\alpha$, but bearing in mind that this analysis is valid for any species. Then, equation \eqref{eq:SolutionDistributionFunctionSph} becomes

\begin{align}
    f(t,v,\mu) = \frac{n_0}{\pi^{3/2}v_{th}^3}\exp\left(-\frac{v^2}{v_{th}^2} \left(\frac{n}{n_0}\right)^{-2/3}\left\{ \left(\frac{R}{R_0}\right)^{-1} + \mu^2\left[ \left(\frac{R}{R_0}\right)^2 - \left(\frac{R}{R_0}\right)^{-1} \right] \right\} \right),
\end{align}

where $R/R_0 = (n/n_0)^{-2/3}(B/B_0)$. Notably, we can do a coordinate transformation from $(v,\mu)$ to $(v_{\perp},v_{\parallel})$ and recover the well-known Bi-Maxwellian distribution

\begin{align}
    f(t,v,\mu) &= \frac{n_0}{\pi^{3/2}v_{th}^3}\exp\left( -\frac{v_{\perp}^2}{v_{th}^2(B/B_0)} - \frac{v_{\parallel}^2}{v_{th}^2(n/n_0)^{2}(B/B_0)^{-2}} \right)\\
    f(t,v,\mu) &= \frac{n_0}{\pi^{3/2}v_{th}^3}\exp\left( -\frac{mv_{\perp}^2}{k_BT_{\perp}(t)} - \frac{mv_{\parallel}^2}{k_BT_{\parallel}(t)} \right), \\
\end{align}

where in the last step we replaced $T_{\perp} = T_0(B/B_0)$ and $T_{\parallel} = T_0(n/n_0)^{2}(B/B_0)^{-2}$, as we naturally recover the expected, nonrelativistic CGL evolution for $T_{\perp}(t)$ and $T_{\parallel}(t)$ \citep{CGL1956}. Indeed, if we take the $P_{\perp}$ and $P_{\parallel}$ nonrelativistic moments of the distribution function \eqref{eq:SolutionDistributionFunctionSph}, we also obtain the same result,

\begin{align}
    \label{eq:PperpclassicCGL}
    P_{\perp} &= m\int \frac{v^2(1 - \mu^2)}{2} f(t,v,\mu) d^3 v = P_{0} \left(\frac{n}{n_0}\right)\left(\frac{B}{B_0}\right)\\
    P_{\parallel} &= m\int v^2\mu^2 f(t,v,\mu) d^3 v = P_{0} \left(\frac{n}{n_0}\right)^{3}\left(\frac{B}{B_0}\right)^{-2},
    \label{eq:PparclassicCGL}
\end{align}

where $P_0 = n_0k_BT_0$, and the same can be obtained for the density $n(t)$. Given that our distribution functions are gyrotropic and neglecting finite Larmor radius (FLR) effects, we can describe the pressure tensor by perpendicular and parallel scalars, $P_{\perp}$ and $P_{\parallel}$. Then, we have recovered all the well known results in the nonrelativistic regime\footnote{It should be noted that the classic CGL equations can also be obtained by directly taking the appropriate nonrelativistic moments to the drift kinetic equation \eqref{eq:DriftKineticEquationSph} itself. This procedure however, cannot be naturally extended to the relativistic regime, given the modifications in the calculation of the moments of the distribution function in the relativistic case (cf. eqns. \eqref{eq:RelativisticMomentIntegral}).}. We will now see that when considering a Maxwell-Jüttner distribution, we obtain departures from the classic, nonrelativistic CGL, and these new relations constitute the appropriate extension of the double-adiabatic equations to the relativistic and ultrarelativistic regimes for the initial distribution considered.

\subsection{Ultrarelativistic case: Maxwell-Jüttner distribution}\label{sec:Ultrarelativistic}

Let us start our relativistic extension of the double-adiabatic equations with the ultrarelativistic case $\gamma = \sqrt{1 + (p/mc)^2} \gg 1$, as it is analytically more tractable.

Consider now the case when the initial distribution function $f_{0}$ is an ultrarelativistic Maxwell-Jüttner distribution

\begin{align}
    f_{0}(p) = \frac{n_0}{8\pi p_T^3}\exp\left(-\frac{p}{p_T}\right),
\end{align}

where $p_T \equiv k_BT_0/c$ is the thermal momentum. Then, the distribution function \eqref{eq:SolutionDistributionFunctionSph} becomes

\begin{align}
    f(t,p,\mu) = \frac{n_0}{8\pi p_T^3}\exp\left( -\frac{p}{p_T}\left(\frac{n}{n_0}\right)^{-1/3}\left(\frac{R}{R_0}\right)^{-1/2}\left\{ 1 + \mu^2\left[ \left(\frac{R}{R_0}\right)^3 - 1 \right] \right\}^{1/2} \right)
    \label{eq:SolutionDistributionFunctionUR}
\end{align}

where $R(t) \equiv n(t)^{-2/3}B(t)$. The evolution of the perpendicular and parallel components of the pressure tensor are obtained by taking the corresponding moments of the distribution \eqref{eq:SolutionDistributionFunctionUR}. Recalling that (e.g. \citet{RezzollaZanotti2013})

\begin{align}
    n &= \int f(t,p,\mu) d^3p,
\end{align}

and

\begin{align}
    T^{\mu\nu} &= c\int p^{\mu}p^{\nu} f \frac{d^3p}{p^0} \rightarrow c\int p^{\mu}p^{\nu} f \frac{d^3p}{p},
    \label{eq:RelativisticMomentIntegral}
\end{align}

where $p^{\mu}$, $p^{\nu}$ are the $\mu$ and $\nu$ components of the four-momentum, and the last expression in \eqref{eq:RelativisticMomentIntegral} is obtained by taking the ultrarelativistic limit $p^0 = \gamma mc \rightarrow p$. We then obtain, for the perpendicular pressure

\begin{align}
    P_{\perp} &= c\int \frac{p^2(1 - \mu^2)}{2} f(t,p,\mu) \frac{d^3 p}{p}\\
    P_{\perp} &= \frac{3}{4}P_0 \left(\frac{n}{n_0}\right)^{4/3}\left(\frac{R}{R_0}\right)^2\left\{ \frac{(R/R_0)^3-2}{[(R/R_0)^3-1]^{3/2}}\arctan\left(\sqrt{(R/R_0)^3-1}\right) + \frac{1}{(R/R_0)^3-1} \right\}
    \label{eq:Pperp_UR}
\end{align}
\normalsize

for $R>1$, and 

\begin{align}
    P_{\perp} &= \frac{3}{4}P_0\left(\frac{n}{n_0}\right)^{4/3}\left(\frac{R}{R_0}\right)^2 \left\{ \frac{2 - (R/R_0)^3}{[1-(R/R_0)^3]^{3/2}}\text{arctanh}\left(\sqrt{1 - (R/R_0)^3}\right) - \frac{1}{1 - (R/R_0)^3} \right\},
\end{align}

\normalsize

for $R<1$. For the parallel pressure,

\begin{align}
    P_{\parallel} &= c\int p^2\mu^2 f \frac{d^3 p}{p}\\
    P_{\parallel} &= \frac{3}{2}P_0 \left(\frac{n}{n_0}\right)^{4/3}\left(\frac{R}{R_0}\right)^2\left\{ \frac{\arctan\left( \sqrt{(R/R_0)^3 - 1} \right)}{[(R/R_0)^3-1]^{3/2}} - \frac{1}{(R/R_0)^3[(R/R_0)^3-1]}\right\},
    \label{eq:Ppar_UR}
\end{align}

\normalsize

for $R>1$, and 

\normalsize
\begin{align}
    P_{\parallel} &= \frac{3}{2}P_0\left(\frac{n}{n_0}\right)^{4/3}\left(\frac{R}{R_0}\right)^2\left\{\frac{1}{(R/R_0)^3[1-(R/R_0)^3]} - \frac{\text{arctanh}\left(\sqrt{1 - (R/R_0)^3}\right)}{[1 - (R/R_0)^3]^{3/2}} \right\},
    \label{eq:Ppar_UR_decreasing}
\end{align}

for $R<1$. It can be shown that when $B\rightarrow B_0$ and $n\rightarrow n_0$, $P_{\perp} \rightarrow P_0$ and $P_{\parallel} \rightarrow P_0$, effectively recovering the initial isotropic state. An analogous calculation can be done to obtain the expected density evolution.

We can also calculate the internal energy density $U$ in order to compare with the general state equations from \citet{Gedalin1991}

\begin{align}
    U &= c\int p^2 f(t,p,\mu) \frac{d^3p}{p}\\
    U &= \frac{3}{2}P_0\left(\frac{n}{n_0}\right)^{4/3}\left(\frac{R}{R_0}\right)^2\left\{\frac{\arctan\left(\sqrt{(R/R_0)^3-1}\right)}{\sqrt{(R/R_0)^3-1}} + \frac{1}{(R/R_0)^3}\right\}
\end{align}

for $R>1$, and 

\begin{align}
    U &= c\int p^2 f(t,p,\mu) \frac{d^3p}{p}\\
    U &= \frac{3}{2}P_0\left(\frac{n}{n_0}\right)^{4/3}\left(\frac{R}{R_0}\right)^2\left\{ \frac{\text{arctanh}\left(\sqrt{1 - (R/R_0)^3}\right)}{\sqrt{1 - (R/R_0)^3}} + \frac{1}{(R/R_0)^3}   \right\}
\end{align}

for $R<1$ (The two limits smoothly connect at $R=1$ giving the correct value in the isotropic case). After a bit of algebra, it can be shown that $P_{\perp}$ and $P_{\parallel}$ satisfy the general state equations from \citet{Gedalin1991}:

\begin{align}
    P_{\parallel} &= n^2\frac{\partial}{\partial n}\left( \frac{U}{n} \right)\\
    P_{\perp}     &= P_{\parallel} + nB\frac{\partial }{\partial B}\left(\frac{U}{n}\right).
\end{align}

Equations \eqref{eq:Pperp_UR}---\eqref{eq:Ppar_UR_decreasing} constitute the extension of the double-adiabatic equations to the ultrarelativistic regime. We will see in section \ref{sec:simulationresults} that these relations correctly describe the evolution of $P_{\perp}$ and $P_{\parallel}$ in PIC simulations of shearing and compressing boxes.

\subsection{Relativistic case: Maxwell-Jüttner distribution}
\label{sec:Relativistic}

Let us finally consider the full Maxwell-Jüttner distribution, valid for fully non-relativistic, moderately relativistic, and ultrarelativistic temperatures

\begin{align}
    f_0(p) = \frac{n_0}{4\pi m^3c^3 \theta K_2(1/\theta)}\exp\left(-\frac{1}{\theta}\sqrt{1 + \frac{p^2}{m^2c^2}}\right),
\end{align}

where $\theta \equiv k_BT_0/mc^2$ is the normalized temperature and $K_2(x)$ is the modified Bessel function of the second kind. Equation \eqref{eq:SolutionDistributionFunctionSph} then becomes

\small
\begin{align}
    f(t,p,\mu) = \frac{n_0}{4\pi m^3c^3 \theta K_2(1/\theta)}\exp\left(-\frac{1}{\theta}\sqrt{1 + \frac{p^2}{m^2c^2}\left(\frac{n}{n_0}\right)^{-2/3}\left(\frac{R}{R_0}\right)^{-1}\left\{ 1 + \mu^2\left[ \left(\frac{R}{R_0}\right)^3 - 1 \right] \right\}}\right),
    \label{eq:SolutionDistributionFunction_MJ}
\end{align}
\normalsize

where $R(t) = n(t)^{-2/3}B(t)$. In cylindrical coordinates in momentum, equation \eqref{eq:SolutionDistributionFunction_MJ} reads

\begin{align}
    f(t,p_{\perp},p_{\parallel}) = \frac{n_0}{4\pi m^3c^3 \theta K_2(1/\theta)}\exp\left(-\frac{1}{\theta}\sqrt{1 + \frac{p_{\perp}^2/m^2c^2}{(B/B_0)} + \frac{p_{\parallel}^2/m^2c^2}{(n/n_0)^{2}(B/B_0)^{-2}}}\right).
    \label{eq:SolutionDistributionFunction_MJ_Cyl}
\end{align}

We see that initially, when $n=n_0$, $B=B_0$, the Maxwell-Jüttner distribution is recovered. At any later time, as long that $n(t)$ or $B(t)$ evolve (either growing or decreasing), a finite pressure anisotropy will develop, as we will see below. Therefore, equations \eqref{eq:SolutionDistributionFunction_MJ}, \eqref{eq:SolutionDistributionFunction_MJ_Cyl} constitute a natural extension of the Maxwell-Jüttner to a relativistic anisotropic distribution, analogous to the Bi-Maxwellian for the nonrelativistic Maxwell-Boltzmann.

We note that, in this case, equation \eqref{eq:SolutionDistributionFunction_MJ_Cyl} does not provide a natural way of obtaining expressions for $T_{\perp}$ and $T_{\parallel}$ directly from its functional form, as the normalized temperature $\theta$ remains outside the square root. 

By using the relativistic moment equations for the perpendicular and parallel components of the pressure tensor in a well-magnetized plasma, (c.f. eqn. \eqref{eq:RelativisticMomentIntegral}), we can obtain evolution equations for $P_{\perp}$ and $P_{\parallel}$ in the relativistic case. For the perpendicular pressure

\begin{align}
    P_{\perp} &= c\int \frac{p^2(1 - \mu^2)}{2} f \frac{d^3p}{p^0}\\
    P_{\perp} &= \frac{n_0c}{2 m^4c^4 \theta K_2(1/\theta)}\int \frac{1}{2}\frac{p^4(1 - \mu^2)}{\sqrt{1 + p^2/m^2c^2}} \exp\left( -\frac{1}{\theta}\sqrt{1 + A_{\mu}\frac{p^2}{m^2c^2} } \right) dp d\mu, 
    \label{eq:Pperp_MJ}
\end{align}

and for the parallel pressure,

\begin{align}
    P_{\parallel}  &= c\int p^2\mu^2 f \frac{d^3p}{p^0}\\
     P_{\parallel} &= \frac{n_0c}{2 m^4c^4 \theta K_2(1/\theta)}\int \frac{p^4\mu^2}{\sqrt{1 + p^2/m^2c^2}} \exp\left( -\frac{1}{\theta}\sqrt{1 + A_{\mu}\frac{p^2}{m^2c^2} } \right) dp d\mu,
     \label{eq:Ppar_MJ}
\end{align}

where, for simplicity, we defined $A_{\mu} \equiv  (n/n_0)^{-2/3}(R/R_0)^{-1}\left\{ 1 + \mu^2\left[ (R/R_0)^3 - 1 \right] \right\}$. Note that, in the integrand of both equations \eqref{eq:Pperp_MJ} and \eqref{eq:Ppar_MJ}, the asymmetry between the square root expressions inside the exponential term (containing the $A_{\mu}$ term) and outside of it in the denominator makes the integral harder to solve analytically. Nevertheless, equations \eqref{eq:Pperp_MJ} and \eqref{eq:Ppar_MJ} provide   a valid extension of the double-adiabatic equations to the relativistic regime for any temperature $\theta$. By numerically integrating equations \eqref{eq:Pperp_MJ} and \eqref{eq:Ppar_MJ}, we confirm our results comparing them with the evolution of $P_{\perp}$ and $P_{\parallel}$ in PIC simulations of shearing and expanding boxes. Similarly, we confirm that equations \eqref{eq:Pperp_MJ} and \eqref{eq:Ppar_MJ} also work in the ultrarelativistic regime $\theta \gg 1$, validating the analytical expressions we obtain in that case (c.f. eqn. \eqref{eq:Pperp_UR}, \eqref{eq:Ppar_UR}).

\section{Simulation Setup}\label{sec:simulationsetup}

To validate our results, we use the relativistic, 2.5 dimensional, fully kinetic PIC code TRISTAN-MP \citep{Buneman1993,Spitkovsky2005} to simulate a collisionless plasma made of singly charged ions and electrons subject to two types of external driving: shearing and compression. In both types of driving, ions and electrons in our simulations are initialized with Maxwell-Jüttner distributions with equal initial temperatures $T_i^{\text{init}} = T_e^{\text{init}}$, and $k_BT_i^{\text{init}}/m_i c^2 = 0.2$ and $30$. The physical parameters of our simulations are the initial temperature of ions and electrons, the initial plasma beta $\beta_i^{\text{init}}$, the mass ratio between ions and electrons $m_i/m_e$, and the scale separation parameter, defined as the ratio between the nonrelativistic cyclotron frequency of the ions and the shearing frequency $\Omega_i^{\text{init}}/s$ for the shearing simulations, and the ratio between the nonrelativistic cyclotron frequency of the ions and the compressing frequency $\Omega_i^{\text{init}}/q$ for the compressing simulations. 
We chose $\Omega_i^{\text{init}}/s$ and $\Omega_i^{\text{init}}/q$ large enough to ensure adiabatic invariants are well conserved in the simulations. The numerical parameters of our simulations are the number of macroparticles per cell $N_{\text{ppc}}$, the plasma skin depth in terms of grid point spacing, $\left.c\middle/(\omega_{p,e}^2 + \omega_{p,i}^2)^{1/2}\middle/\Delta x\right.$, where $\omega_{p,e}$ and $\omega_{p,i}$ are the electron and ion plasma frequencies, respectively, and the domain size in terms of the initial ion Larmor radius $L/R_{L,i}$ where $R_{L,i} = v_{th,i}/\Omega_i^{\text{init}}$, and $v_{th,i}^2 = k_BT^{\text{init}}_i/m_i$. The physical and numerical parameters of the shearing and compressing simulations presented here are listed in table \ref{table:ShearingSims} and \ref{table:CompressingSims}.

The shearing and compression motions will generate $\dot{B}/B \neq 0$ and $\dot{n}/n \neq 0$, therefore creating a pressure anisotropy in the plasma, and $P_{\perp,\alpha}$, $P_{\parallel,\alpha}$ ($\alpha=i,e$) will evolve independently. This adiabatic evolution will, in general, continue until the pressure anisotropy surpasses some instability threshold, usually scaling with $1/\beta_{\alpha}$ to some power, at which point the pressure anisotropy will act as a free energy source to make the instability grow. 

The specific instability excited will depend on the physical parameters of the plasma in each simulation. Some examples of pressure anisotropy driven instabilities are ion mirror \citep{Chandra1958,rudakov1959,vedenov1959,Barnes1966,Hasegawa1969,SouthwoodKivelson1993,KivelsonSouthwood1996,Pokhotelov2002,Pokhotelov2004}, ion firehose \citep{Chandra1958, Parker1958,VedenovSagdeev1961,Yoon1993,Gary1998,HellingerMatsumoto2000, Bott2024}, ion cyclotron \citep{SagdeevShafranov1960, Gary1992, Gary1993, Lopez2016}, electron mirror \citep{HellingerStverak2018}, electron firehose \citep{LiHabbal2000,GaryNishimura2003,Lopez2022}, and electron whistler instabilities \citep{KennelPetschek1966,Gary1996}, among others (see e.g. \citet{Bott2024} for a thorough review). Many of these instabilities have been studied in great detail using PIC simulations (e.g. \citet{Gary2011, Kunz2014, Sironi2015, Riquelme2015, Riquelme2016, Riquelme2018, Innocenti2019, Bott2021, Zhdankin2023, Ley2024}).

In this work, we are interested in describing the adiabatic evolution of $P_{\perp,j}$ and $P_{\parallel,j}$, before any instability is excited. Therefore, we will avoid capturing the excitation of any instability in the simulations by either stopping the simulation before excitation (for the shearing case), or disabling the fluctuating field evolution by not depositing particle current (for the compressing case). We show that our results are independent of the mass ratio $m_i/m_e$, and that they are applicable in the three initial plasma beta cases we tried: $\beta_{i}^{\text{init}}=0.05, 0.5, $ and $5$.

\begin{table}
\begin{center}
\caption{The physical and numerical parameters of the shearing simulations. The physical parameters are the initial ion plasma beta $\beta_i^{\text{init}} = 8\pi n_ik_BT_{i}^{\text{init}}/B^2$, the ion to electron mass ratio $m_i/m_e$, the initial ion temperature normalized to the rest mass energy $k_BT_i^{\text{init}}/m_i c^2$, and the scale separation parameter $\Omega_i/s$. The numerical parameters are the number of macroparticles per cell $N_{\text{ppc}} = 300$ in all the simulations, the plasma skin depth in units of grid point spacing $d/\Delta x \equiv \left.c\middle/(\omega_{p,e}^2 + \omega_{p,i}^2)^{1/2}\middle/\Delta x\right.$, and the domain size in units of initial ion Larmor radius $L/R_{L,i}$.}
\label{table:ShearingSims}
\def~{\hphantom{0}}
\begin{tabular}{lccccccc}
\hline
Runs                   & $\beta_i^{\text{init}}$ & $m_i/m_e$ & $k_BT^{\text{init}}/m_ic^2$ & $d/\Delta x$ & $L/R_{L,i}$ & $R_{L,i}/\Delta x$ & $\Omega_{i}/s$ \\
Shb0.5m1836d30wcis3200 & 0.5                     & 1836      & 30                         & 1                                          & 48    & 22      & 3200             \\
Shb0.05m8d30wcis3200   & 0.05                    & 8         & 30                         & 10                                         & 54    & 5     & 3200             \\
Shb0.5m8d0.2wcis800    & 0.5                     & 8         & 0.2                        & 7                                          & 49    & 11      & 800              \\
Shb0.5m8d30wcis3200    & 0.5                     & 8         & 30                         & 7                                          & 97    & 11      & 3200             \\
Shb0.5m64d30wcis3200   & 0.5                     & 64        & 30                         & 3                                          & 85    & 12      & 3200             \\
Shb5m8d30wcis3200      & 5                       & 8         & 30                         & 5                                          & 43    & 24      & 3200            
\end{tabular}
\end{center}
\end{table}

\subsection{Shearing Simulations Setup}
For the shearing simulations, a periodic velocity shear $\textbf{v} = -sx\hat{y}$ is imposed in the domain, where $s$ is the shearing frequency  and $x$ the distance along the $\hat{x}$ axis. The magnetic field initially points along the $x$-axis, $\textbf{B} = B_0\hat{x}$. The PIC system of equations is solved using shearing coordinates \citep{Riquelme2012}. By flux conservation, the action of the shear continuously amplifies the magnetic field strength such that the magnitude of $\textbf{B}$ evolves as $B(t) = B_0\sqrt{1 + s^2t^2}$, where $B_0$ is the initial field strength, whereas the density remains constant, $n=n_0$, where $n_0$ is the initial density. Therefore we have a contribution $\dot{B}/B \neq 0$ and $\dot{n}/n = 0$. In this case, the magnetic field amplification will continuously drive an ion and electron pressure anisotropy, $P_{\perp,i} > P_{\parallel,i}$ and $P_{\perp,e} > P_{\parallel,e}$, respectively, and we will see that the evolution of the perpendicular and parallel pressures will correctly be described by the relations presented in sections \ref{sec:Ultrarelativistic} and \ref{sec:Relativistic}, for mildy relativistic and ultrarelativistic initial temperatures.

\subsection{Compressing Simulations Setup}
For the compressing simulations, a global compression motion is imposed in the domain, and the PIC system of equations is solved in comoving coordinates \citep{Sironi2015}. To achieve this, a transformation from the laboratory frame $(t_{\text{lab}},\textbf{x}_{\text{lab}})$ to the comoving coordinate $(t',\textbf{x}')$ is performed via $\textbf{x}_{\text{lab}} = \textbf{L}\textbf{x}'$, where

\begin{align}
    \textbf{L} = 
    \begin{pmatrix}
    a_x(t) & 0      & 0 \\
    0      & a_y(t) & 0 \\
    0      &   0    & a_z(t)
    \end{pmatrix},
\end{align}

and the differential transformation law reads:

\begin{align}
    d\textbf{x}_{\text{lab}} = \textbf{L}d\textbf{x}' + \dot{\textbf{L}}\textbf{x}'dt'.
\end{align}

\begin{table}
\begin{center}
\caption{The physical and numerical parameters of the compressing simulations. The physical parameters are the initial ion plasma beta $\beta_i^{\text{init}} = 8\pi n_ik_BT_{i}^{\text{init}}/B^2$, the ion to electron mass ratio $m_i/m_e$, the initial ion temperature normalized to the rest mass energy $k_BT_i^{\text{init}}/m_i c^2$, and the scale separation parameter $\Omega_i/q$. The numerical parameters are the number of macroparticles per cell $N_{\text{ppc}} = 300$ in all the simulations, the plasma skin depth in units of grid point spacing $d/\Delta x \equiv \left.c\middle/(\omega_{p,e}^2 + \omega_{p,i}^2)^{1/2}\middle/\Delta x\right.$, and the domain size in units of initial ion Larmor radius $L/R_{L,i}$.}
\label{table:CompressingSims}
\def~{\hphantom{0}}
\begin{tabular}{lccccccc}
\hline
Runs                     & $\beta_i^{\text{init}}$ & $m_i/m_e$ & $k_BT^{\text{init}}/mic^2$ & $d/\Delta x$ & $L/R_{L,i}$ & $R_{L,i}/\Delta x$ & $\omega_{c,i}/q$ \\
Compb0.5m1836d30wcis3200 & 0.5                     & 1836      & 30                         & 1                                          & 12    & 21      & 3200             \\
Compb0.05m8d30wcis3200   & 0.05                    & 8         & 30                         & 10                                         & 54    & 5      & 3200             \\
Compb0.5m8d0.2wcis800    & 0.5                     & 8         & 0.2                        & 7                                          & 24    & 11      & 800              \\
Compb0.5m8d30wcis3200    & 0.5                     & 8         & 30                         & 7                                          & 24    & 11      & 3200             \\
Compb0.5m64d30wcis3200   & 0.5                     & 64        & 30                         & 3                                          & 21    & 12      & 3200             \\
Compb5m8d30wcis3200      & 5                       & 8         & 30                         & 3                                          & 18    & 24      & 3200            
\end{tabular}
\end{center}
\end{table}

In our 2D domain, the magnetic field is aligned in the $x$ direction, and the compression is performed along both $y$ and $z$ axes, i.e. along the perpendicular directions with respect to the direction of the magnetic field $\textbf{B}$. For this, we choose the following scale factors:

\begin{align}
    a_x(t) &= 1, \\
    a_y(t) &= a_z(t) = \frac{1}{1 + qt},
\end{align}

where $q$ is the compression rate, which can be controlled for each simulation. By flux freezing, both density and  magnetic field strength get amplified, evolving as:

\begin{align}
    n(t) &= n_0(1 + qt)^2\\
    B(t) &= B_0(1 + qt)^2,
\end{align}

where $n_0$ is the initial density, and $B_0$ is the initial field strength. Therefore, in this case we have $\dot{n}/n \neq 0$ and $\dot{B}/B \neq 0$. The magnetic field and density amplification will also drive a continuous ion and electron pressure anisotropy, $P_{\perp,i} > P_{\parallel,i}$ and $P_{\perp,e} > P_{\parallel,e}$, respectively.

It is important to note that in the compressing simulations the parallel component of the momentum of both ions and electrons remain unchanged during the compression, $p_{\parallel} = p_{\parallel,0}$. Therefore, $dp_{\parallel}/dt = 0$ in the drift kinetic equation \eqref{eq:DKE_raw}. We can solve this special case following the same procedure as in section \ref{sec:theoreticalBasis}, and the evolution of $P_{\perp}$ and $P_{\parallel}$ for this case will be presented in section \ref{sec:simulationresults_compressing} along with the simulation results, also showing good agreement.

\section{Simulation Results}\label{sec:simulationresults}

\subsection{Shearing Simulations}\label{sec:simulationresults_shearing}

Shearing simulations generate a growing magnetic field $B(t)/B_0 = \sqrt{1 + s^2t^2}$ and a constant density, $n/n_0=1$. We can then replace these relations in the relativistic double-adiabatic equations \eqref{eq:Pperp_UR},\eqref{eq:Ppar_UR} or \eqref{eq:Pperp_MJ},\eqref{eq:Ppar_MJ} depending on the initial temperature, in order to compare with simulations. 

\subsubsection{Initial $k_BT_i^{\text{init}}/m_ic^2= 0.2$}

\begin{figure}
    \centering
    \includegraphics[width=0.85\linewidth]{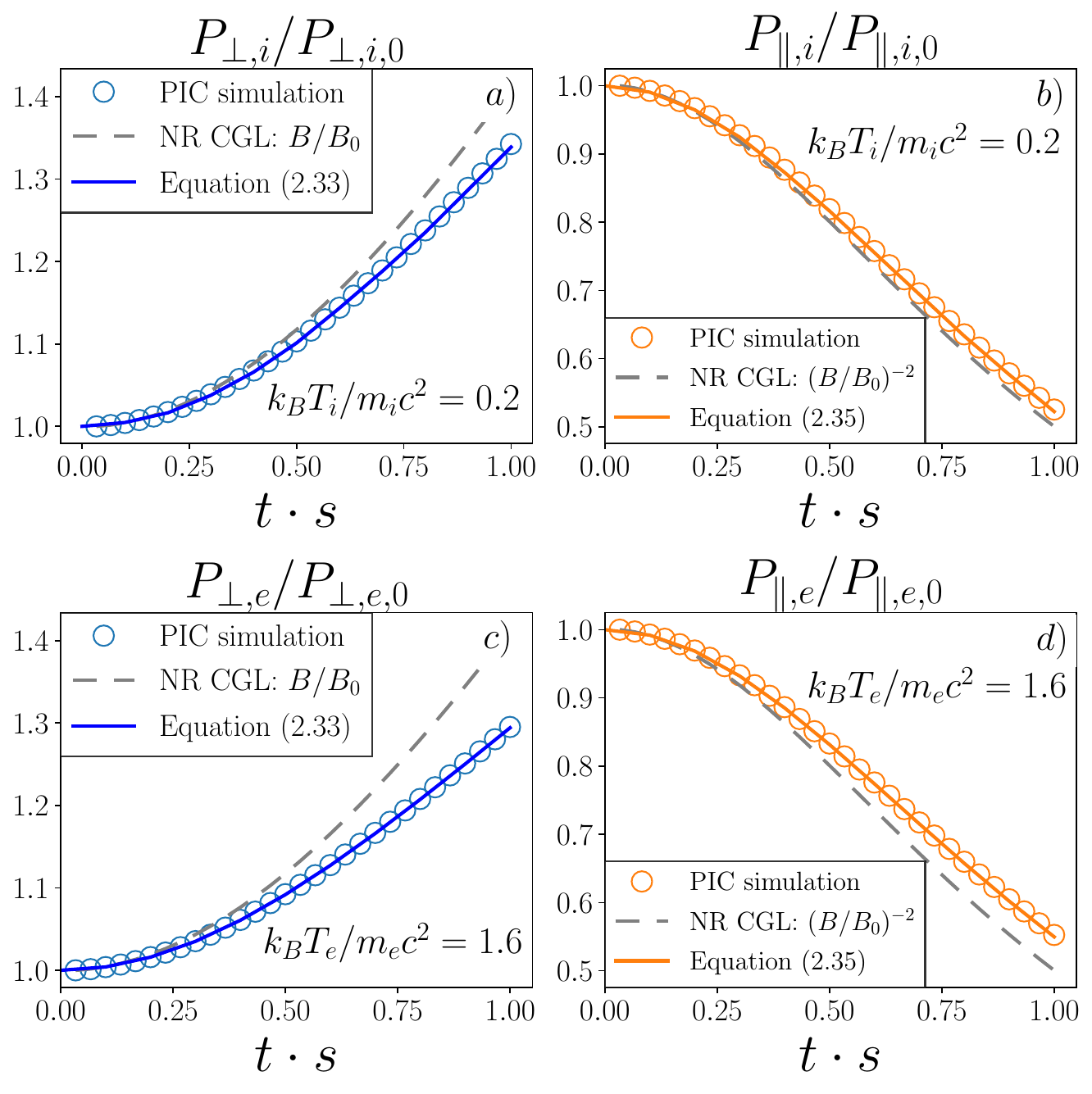}
    \caption{Panel $a$: The evolution of the ion perpendicular pressure for run Shb0.5m8d0.2wcis800 (open blue circles), for initial $k_BT_i^{\text{init}}/m_ic^2=0.2$. The solid blue line shows the evolution of $P_{\perp}$ according to eqn. \eqref{eq:Pperp_MJ}, numerically integrated. Panel $b$: The evolution of the ion parallel pressure (open orange circles). The solid orange line shows the evolution of $P_{\parallel}$ according to eqn. \eqref{eq:Ppar_MJ}, numerically integrated. Panel $c$: The evolution of the electron perpendicular pressure (open blue circles), for initial $k_BT_e^{\text{init}}/m_ec^2=1.6$ (given a mass ratio $m_i/m_e = 8$). The solid blue line shows the evolution of $P_{\perp}$ according to eqn. \eqref{eq:Pperp_MJ}, numerically integrated. Panel $d$: The evolution of the electron parallel pressure (open orange circles). The solid orange line shows the evolution of $P_{\parallel}$ according to eqn. \eqref{eq:Ppar_MJ}, numerically integrated. In panels $a$ and $c$, the nonrelativistic CGL evolution for $P_{\perp}$ is shown in dashed gray line. In panels $b$ and $d$, the nonrelativistic CGL evolution for $P_{\parallel}$ is shown in dashed gray line.}
    \label{fig:Shear_delgam0.2_ion}
\end{figure}

The evolution of the ion perpendicular pressure $P_{\perp,i}$ and parallel pressure $P_{\parallel,i}$ as well as the evolution of the electron perpendicular and parallel pressure $P_{\perp,e}$, $P_{\parallel,e}$ are shown in figure \ref{fig:Shear_delgam0.2_ion} for run Shb0.5m8d0.2wcis800.

For this mildly relativistic temperature, we can already see that $P_{\perp,i}$ (open blue circles in panel $a$) departs from the nonrelativistic CGL prediction for a pure shearing, $B/B_0$ (dashed gray line in panel $a$, cf. eqn. \eqref{eq:PperpclassicCGL}). In the parallel pressure (open orange circles in panel $b$), we also see an analogous departure from the CGL prediction $(B/B_0)^{-2}$, although it is less pronounced than in the perpendicular case.

Our relativistic extension for the evolution of $P_{\perp}$ is shown in solid blue line in panel $a$ of figure \ref{fig:Shear_delgam0.2_ion}. This line is obtained by numerically integrating equation \eqref{eq:Pperp_MJ} for $\theta=0.2$. We can see that it consistently follows the simulation results for $P_{\perp,i}$. Similarly, the relativistic extension for the evolution of $P_{\parallel}$, obtained by numerically integrating equation \eqref{eq:Ppar_MJ} for $\theta=0.2$ and shown in solid orange line in panel $b)$, also shows very good agreement with the evolution of $P_{\parallel,i}$. We can see that our derivations outlined in section \ref{sec:Relativistic} correctly describe the behavior of $P_{\perp}$ and $P_{\parallel}$ for this mildly relativistic temperature.

We see an analogous behavior for electrons. Given that $m_i/m_e=8$, we initially have $k_BT_e^{\text{init}}/m_ec^2 = k_BT_i^{\text{init}}/m_ic^2 (m_i/m_e) = 1.6$. By numerically integrating equations \eqref{eq:Pperp_MJ}, \eqref{eq:Ppar_MJ} for $\theta=1.6$, we also obtain an evolution of $P_{\perp}$ and $P_{\parallel}$ that is consistent with our PIC results. Indeed, we can see the evolution of $P_{\perp,e}$ (open blue circles, panel $c)$) and $P_{\parallel,e}$ (open orange cirles, panel $d)$) for run Shb0.5m8d0.2wcis800. Equations \eqref{eq:Pperp_MJ} (solid blue line in panel $c)$) and \eqref{eq:Ppar_MJ} (solid orange line in panel $d)$) correctly describe the evolution of $P_{\perp,e}$ and $P_{\parallel,e}$, respectively. In this case, the departure from the nonrelativistic CGL evolution is more pronounced in both perpendicular and parallel pressures, as the initial temperature of electrons is higher by a factor $m_i/m_e=8$.

\subsubsection{Initial $k_BT_i^{\text{init}}/m_ic^2 = 30$}

\begin{figure}
    \centering
    \includegraphics[width=0.85\linewidth]{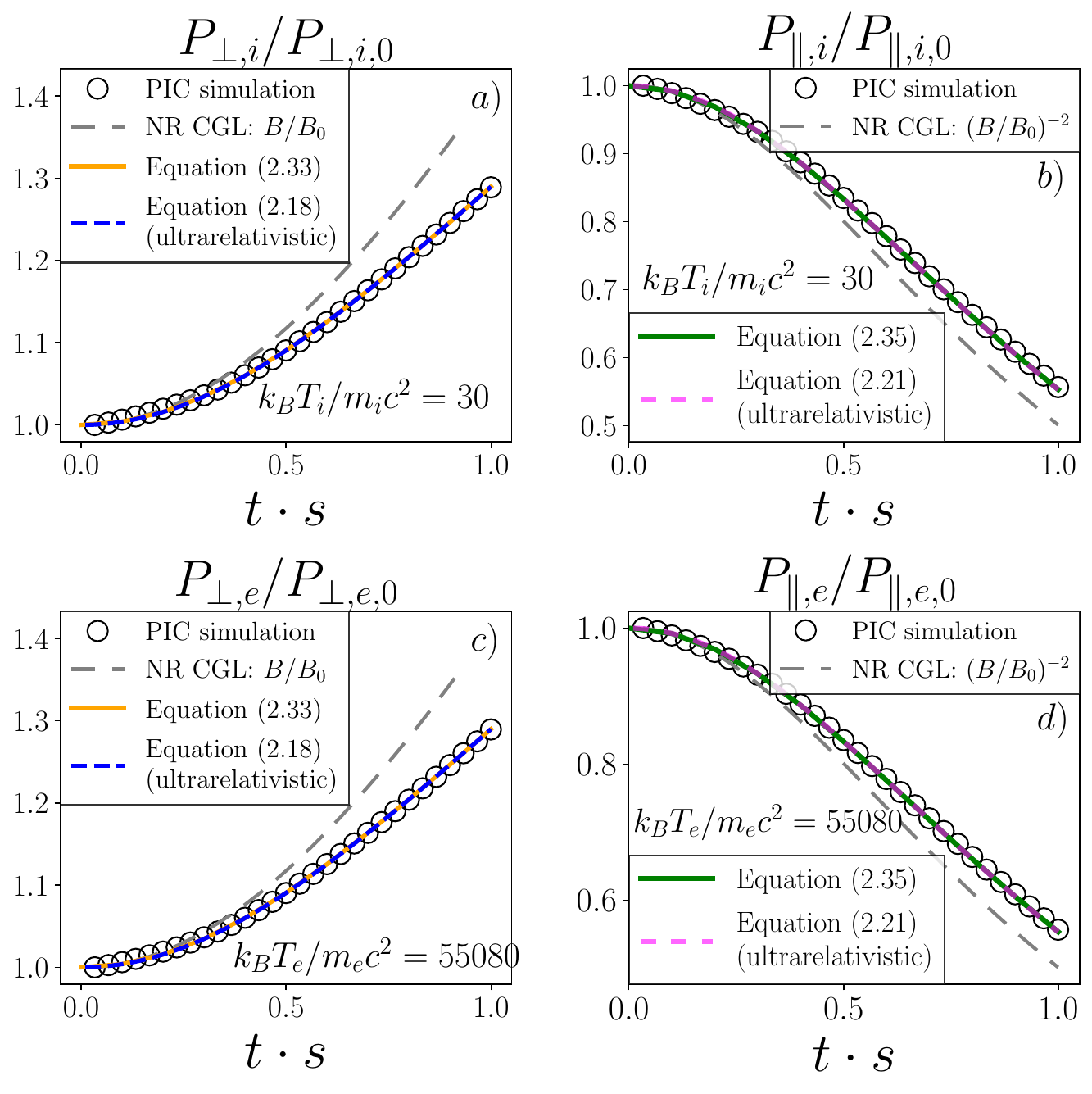}
    \caption{Panel $a$: The evolution of the ion perpendicular pressure for run Shb0.5m1836d30wcis3200 (open black circles), for initial $k_BT_i^{\text{init}}/m_ic^2=30$. The dashed blue line shows the evolution of $P_{\perp}$ according to eqn. \eqref{eq:Pperp_UR}. The solid orange line shows the evolution of $P_{\perp}$ according to eqn. \eqref{eq:Pperp_MJ}, numerically integrated. Panel $b$: The evolution of the ion parallel pressure (open black circles). The dashed magenta line shows the evolution of $P_{\parallel}$ according to eqn. \eqref{eq:Ppar_UR}. The solid green line shows the evolution of $P_{\parallel}$ according to eqn. \eqref{eq:Ppar_MJ}, numerically integrated. Panel $c$: The evolution of the electron perpendicular pressure (open black circles), for initial $k_BT_e^{\text{init}}/m_ec^2=55080$. The dashed blue line shows the evolution of $P_{\perp}$ according to eqn. \eqref{eq:Pperp_UR}. The solid orange line shows the evolution of $P_{\perp}$ according to eqn. \eqref{eq:Pperp_MJ}, numerically integrated. Panel $d$: The evolution of the electron parallel pressure (open black circles). The dashed magenta line shows the evolution of $P_{\parallel}$ according to eqn. \eqref{eq:Ppar_UR}. The solid green line shows the evolution of $P_{\parallel}$ according to eqn. \eqref{eq:Ppar_MJ}, numerically integrated. In panels $a$ and $c$, the nonrelativistic CGL evolution for $P_{\perp}$ is shown in dashed gray line. In panels $b$ and $d$, the nonrelativistic CGL evolution for $P_{\parallel}$ is shown in dashed gray line.}
    \label{fig:Shear_delgam30_ion}
\end{figure}

The evolution of the ion perpendicular pressure $P_{\perp,i}$ and parallel pressure $P_{\parallel,i}$, as well as the evolution of the electron perpendicular and parallel pressure $P_{\perp,e}$, $P_{\parallel,e}$ are shown in figure \ref{fig:Shear_delgam30_ion} for run Shb0.5m1836d30wcis3200. For initial $k_BT_i^{\text{init}}/m_ic^2 = 30$, we are already in the ultrarelativistic regime, so we can now test our analytic derivation of equations \eqref{eq:Pperp_UR} and \eqref{eq:Ppar_UR}.

For the ion perpendicular pressure $P_{\perp,i}$ (panel $a$, open black circles), we can see that it departs from the nonrelativistic CGL prediction $B/B_0$ (dashed gray line) much further than in the lower initial temperature case. In contrast, we observe that our ultrarelativistic extension for the evolution of $P_{\perp}$, equation \eqref{eq:Pperp_UR} (dashed blue line in panel $a)$), consistently follows the evolution of $P_{\perp,i}$. Our general equation \eqref{eq:Pperp_MJ} (numerically integrated for $\theta=30$, solid orange line) is also shown, and it also consistently follows the evolution of $P_{\perp,i}$. Therefore, our analytical expression for $P_{\perp}$ in the ultrarelativistic case is consistent with our general formula, eqn. \eqref{eq:Pperp_MJ}. 

The evolution of the ion parallel pressure $P_{\parallel,i}$ (open black circles, right panel of fig. \ref{fig:Shear_delgam30_ion}) follows a similar trend. The departure from the nonrelativistic CGL prediction $(B/B_0)^{-2}$ (dashed gray line) is also evident, and our ultrarelativistic extension for $P_{\parallel}$, equation \eqref{eq:Ppar_UR} (solid magenta line in fig. \ref{fig:Shear_delgam30_ion}), is also consistent with the simulation results. We show the general formula \eqref{eq:Ppar_MJ} (numerically integrated for $\theta=30$, solid green line), which also explains very well the evolution of $P_{\parallel,i}$. Therefore, our ultrarelativistic analytical result is also consistent with the general formula \eqref{eq:Ppar_MJ}.

Similar to the previous case, the electron pressure in run Shb0.5m1836d30wcis3200 has an analogous evolution, now with a much higher initial temperature, given the realistic mass ratio $m_i/m_e=1836$. Nevertheless, the ultrarelativistic extension derived in section \ref{sec:Ultrarelativistic} is also valid, and correctly describes the evolution of the electron pressure components.

Indeed, the evolution of the electron perpendicular pressure $P_{\perp,e}$ and parallel pressure $P_{\parallel,e}$ for run Shb0.5m1836d30wcis3200 is shown in figure \ref{fig:Shear_delgam30_ion}. Similar to the ion case, we can see that neither $P_{\perp,e}$ (panel $c)$, open black circles) nor $P_{\parallel,e}$ (panel $d)$, open black circles) follow the respective CGL prediction (dashed gray lines in panels $c)$ and $d)$ of figure \ref{fig:Shear_delgam30_ion}), whereas our ultrarelativistic extensions, equations \eqref{eq:Pperp_UR} (dashed blue line in panel $c)$) and \eqref{eq:Ppar_UR} (dashed magenta line in panel $d)$) correctly describes their evolution. We also include the general equations \eqref{eq:Pperp_MJ} (solid orange line, panel $c)$) and \eqref{eq:Ppar_MJ} (solid green line, panel $d)$) numerically integrated for $\theta=30$, showing good agreement as well.

Our ultrarelativistic extensions of the double-adiabatic evolutions of $P_{\perp}$ (eqn. \eqref{eq:Pperp_UR}) and $P_{\parallel}$ (eqn. \eqref{eq:Ppar_UR}) are demonstrated to agree with PIC simulations with a shearing external driving. Additionally, they do not depend on the initial temperature, provided that it is sufficiently high\footnote{In practice, we see very good agreement  once $k_BT/mc^2 \gtrsim 1$.}, analogous to the nonrelativistic, CGL case, which is also independent of temperature (as long as temperatures are sufficiently low).

\subsection{Compressing Simulations}\label{sec:simulationresults_compressing}

As mentioned in section \ref{sec:simulationsetup}, the compressing simulations leave the parallel momentum unchanged, $p_{\parallel}=p_{\parallel,0}$. The constancy of parallel momentum implies that $dp_{\parallel}/dt = 0$ in the drift kinetic equation \ref{eq:DKE_raw}:

\begin{align}
    \frac{\partial f}{\partial t} + \frac{dp_{\perp}}{dt}\frac{\partial f}{\partial p_{\perp}} = \frac{\partial f}{\partial t} + \frac{\dot{B}}{2B}p_{\perp}\frac{\partial f}{\partial p_{\perp}} = 0.
\end{align}

This just constitutes a special case of the general derivation presented in section \ref{sec:theoreticalBasis}. We note that in this case the drift kinetic equation can also readily be solved by the method of characteristics to obtain

\begin{align}
    f(t,p_{\perp},p_{\parallel}) = f_0\left(\frac{p_{\perp}}{(B/B_0)^{1/2}}, p_{\parallel}\right).
    \label{eq:Solution_DKE_Compressing}
\end{align}

For an ultrarelativistic Maxwell-Jüttner as initial condition, our solution \eqref{eq:Solution_DKE_Compressing} becomes

\begin{align}
    f(t,p_{\perp},p_{\parallel}) = \frac{n_0}{8\pi p_T}\exp\left(-\frac{1}{p_T}\sqrt{\frac{p_{\perp}^2}{B/B_0} + p_{\parallel}^2}\right),
\end{align}

and taking the perpendicular and parallel pressure moments (cf. eqns. \eqref{eq:RelativisticMomentIntegral}),

\begin{align}
    \label{eq:Pperp_compressing_UR}
    P_{\perp} = \frac{3}{4}P_0\left(\frac{B}{B_0}\right)^2\left[ \frac{(B/B_0)-2}{((B/B_0)-1)^{3/2}}\arctan\left(\sqrt{(B/B_0)-1}\right) + \frac{1}{(B/B_0)-1} \right]
\end{align}
\normalsize

for $(B/B_0)>1$, and 

\begin{align}
    P_{\perp} = \frac{3}{4}P_0\left(\frac{B}{B_0}\right)^2\left[ \frac{2 - (B/B_0)}{(1-(B/B_0))^{3/2}}\text{arctanh}\left( \sqrt{1 - (B/B_0)} \right) - \frac{1}{1 - (B/B_0)} \right]
\end{align}

\normalsize

for $(B/B_0)<1$. Similarly for the parallel pressure

\begin{align}
    \label{eq:Ppar_compressing_UR}
    P_{\parallel} = \frac{3}{2}P_0\left[ \frac{(B/B_0)^2}{((B/B_0) - 1)^{3/2}}\arctan\left( \sqrt{(B/B_0)-1} \right) - \frac{(B/B_0)}{(B/B_0)-1} \right]
\end{align}

for $(B/B_0)>1$, and 

\begin{align}
    P_{\parallel} = \frac{3}{2}P_0\left[ \frac{(B/B_0)}{1 - (B/B_0)} - \frac{(B/B_0)^2}{(1 - (B/B_0))^{3/2}}\text{arctanh}\left(\sqrt{1 - (B/B_0)}\right) \right]
\end{align}

for $(B/B_0)<1$. 

For an initial Maxwell-Jüttner, our solution \eqref{eq:Solution_DKE_Compressing} becomes

\begin{align}
    f(t,p_{\perp},p_{\parallel}) = \frac{n_0}{4\pi m^3c^3 \theta K_2(1/\theta)}\exp\left( -\frac{1}{\theta}\sqrt{1 + \frac{p_{\perp}^2/m^2c^2}{(B/B_0)} + p_{\parallel}^2/m^2c^2} \right), 
    \label{eq:Solution_DKE_Compressing_MJ}
\end{align}

and the perpendicular and parallel pressures can also be obtained by numerically integrating the moments:

\small
\begin{align}
    P_{\perp} &= \frac{n_0c}{2m^4c^4\theta K_2(1/\theta)} \int \frac{1}{2}\frac{p_{\perp}^2}{\sqrt{1 + p_{\perp}^2/m^2c^2 + p_{\parallel}^2/m^2c^2}} \exp\left( -\frac{1}{\theta}\sqrt{1 + \frac{p_{\perp}^2/m^2c^2}{(B/B_0)} + p_{\parallel}^2/m^2c^2} \right)p_{\perp}dp_{\perp}dp_{\parallel} ,
    \label{eq:Pperp_Compressing_MJ}
\end{align}
\normalsize

and

\small
\begin{align}
    P_{\parallel} &= \frac{n_0c}{2m^4c^4\theta K_2(1/\theta)} \int \frac{p_{\parallel}^2}{\sqrt{1 + p_{\perp}^2/m^2c^2 + p_{\parallel}^2/m^2c^2}} \exp\left( -\frac{1}{\theta}\sqrt{1 + \frac{p_{\perp}^2/m^2c^2}{(B/B_0)} + p_{\parallel}^2/m^2c^2} \right)p_{\perp}dp_{\perp}dp_{\parallel}.
    \label{eq:Ppar_Compressing_MJ}
\end{align}
\normalsize

We can now use equations \eqref{eq:Pperp_compressing_UR},\eqref{eq:Ppar_compressing_UR}, \eqref{eq:Pperp_Compressing_MJ}, and \eqref{eq:Ppar_Compressing_MJ} to compare with our compressing box simulations. In all the results shown below, we pass from pressures to temperatures using $T_{\perp}=P_{\perp}/(n k_B)$ and $T_{\parallel} = P_{\parallel}/(n k_B)$, where $n(t)=n_0(1 + qt)^2$, as in the relativistic regime it also holds that $P=nk_BT$ (e.g. \citet{RezzollaZanotti2013}).

\subsubsection{Initial $k_BT_i^{\text{init}}/m_i c^2 = 0.2$}

\begin{figure}
    \centering
    \includegraphics[width=0.95\linewidth]{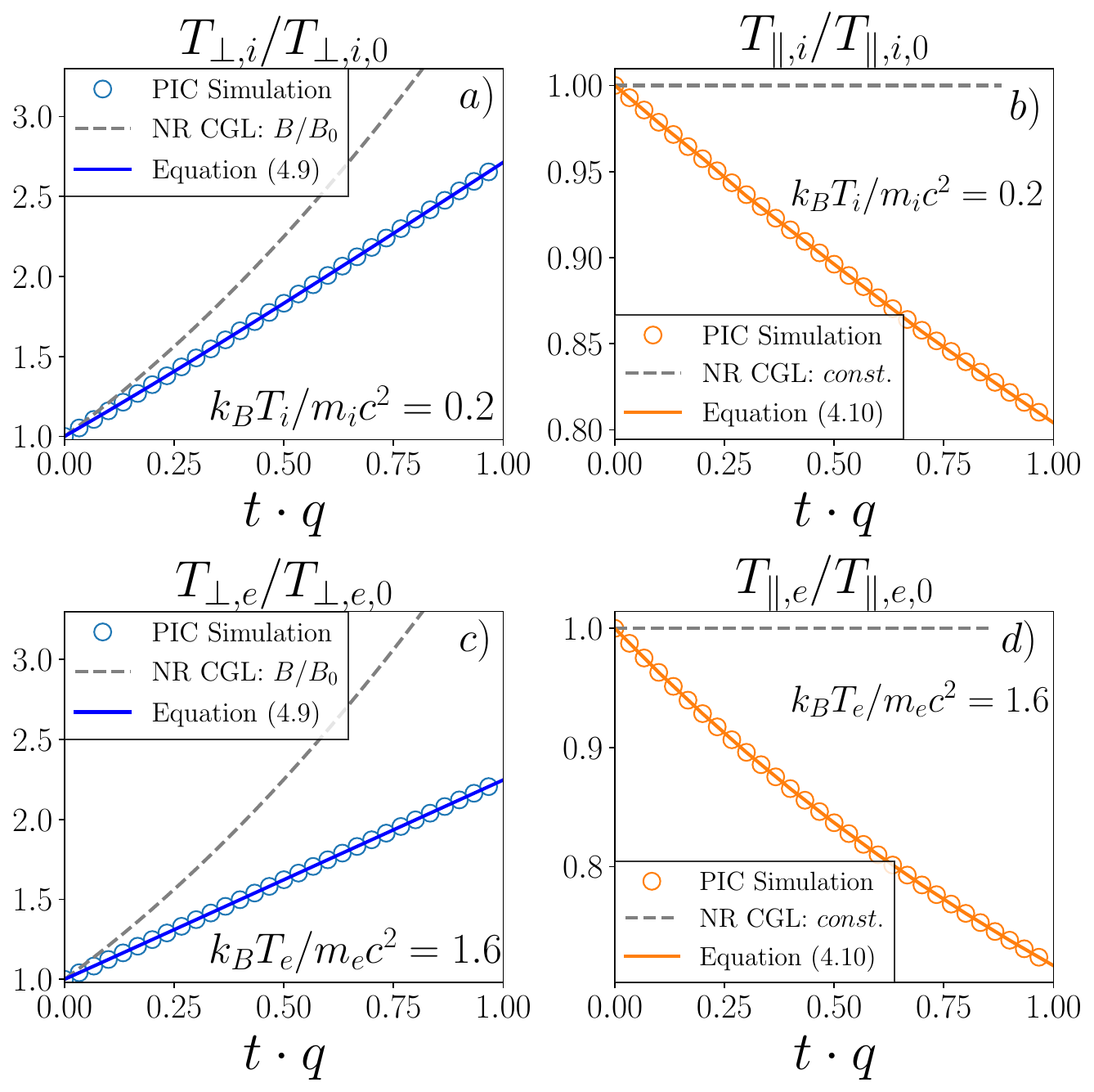}
    \caption{Panel $a$: The evolution of the ion perpendicular temperature for run Compb0.5m8d0.2wcis800 (open blue circles), for initial $k_BT_i^{\text{init}}/m_ic^2=0.2$. The solid blue line shows the evolution of $T_{\perp}$ according to eqn. \eqref{eq:Pperp_Compressing_MJ}, numerically integrated. Panel $b$: The evolution of the ion parallel temperature  (open orange circles). The solid orange line shows the evolution of $T_{\parallel}$ according to eqn. \eqref{eq:Ppar_Compressing_MJ}, numerically integrated. Panel $c$: The evolution of the electron perpendicular temperature (open blue circles), for initial $k_BT_e^{\text{init}}/m_ec^2=1.6$. The solid blue line shows the evolution of $T_{\perp}$ according to eqn. \eqref{eq:Pperp_Compressing_MJ}, numerically integrated. Panel $d$: The evolution of the electron parallel temperature (open orange circles). The solid orange line shows the evolution of $T_{\parallel}$ according to eqn. \eqref{eq:Ppar_Compressing_MJ}, numerically integrated. In panels $a$ and $c$ the nonrelativistic CGL evolution for $P_{\perp}$ is shown in dashed gray line for a compressing motion. In panels $b$ and $d$, the nonrelativistic CGL evolution for $P_{\parallel}$ is shown in dashed gray line for a compressing motion.}
    \label{fig:Comp_delgam0.2_ion}
\end{figure}

The evolution of the ion perpendicular temperature $T_{\perp,i}$ and parallel temperature $T_{\parallel,i}$, and the electron perpendicular and parallel temperatures $T_{\perp,e}$, $T_{\parallel,e}$ are shown in figure \ref{fig:Comp_delgam0.2_ion} for run Compb0.5m8d0.2wcis800.

Similar to the shearing case, we can see that both $T_{\perp,i}$ (open blue circles in panel $a$) and $T_{\parallel,i}$ (open orange circles in panel $b$ ) depart from the nonrelativistic CGL prediction (dashed gray lines). This is especially critical in the parallel case, as the nonrelativistic CGL predicts a constant $T_{\parallel}$. This departure from CGL was also observed in \citet{Tran2023} for the same initial ion temperature.

In contrast, we can see that our relativistic double-adiabatic equations describe very well the temperature evolution in our simulations. We numerically integrated equations \eqref{eq:Pperp_Compressing_MJ} (solid blue line in panel $a$) and \eqref{eq:Ppar_Compressing_MJ} (solid orange line in panel $b$) for $\theta = 0.2$.

The evolution of the electron perpendicular temperature $T_{\perp,e}$ and parallel temperature $T_{\parallel,e}$ for run Compb0.5m8d0.2wcis800 is shown in figure \ref{fig:Comp_delgam0.2_ion}. In this case we also see a very similar behavior, now for $k_BT_e^{\text{init}}=1.6$. Both $T_{\perp,e}$ (open blue circles, panel $c$) and $T_{\parallel,e}$ (open orange circles, panel $d$) depart from their nonrelativistic CGL evolution (dashed gray lines), and are very well described by our relativistic double-adiabatic equations \eqref{eq:Pperp_Compressing_MJ}, \eqref{eq:Ppar_Compressing_MJ}.

\subsubsection{Initial $k_BT_i^{\text{init}}/m_ic^2 = 30$}

\begin{figure}
    \centering
    \includegraphics[width=0.95\linewidth]{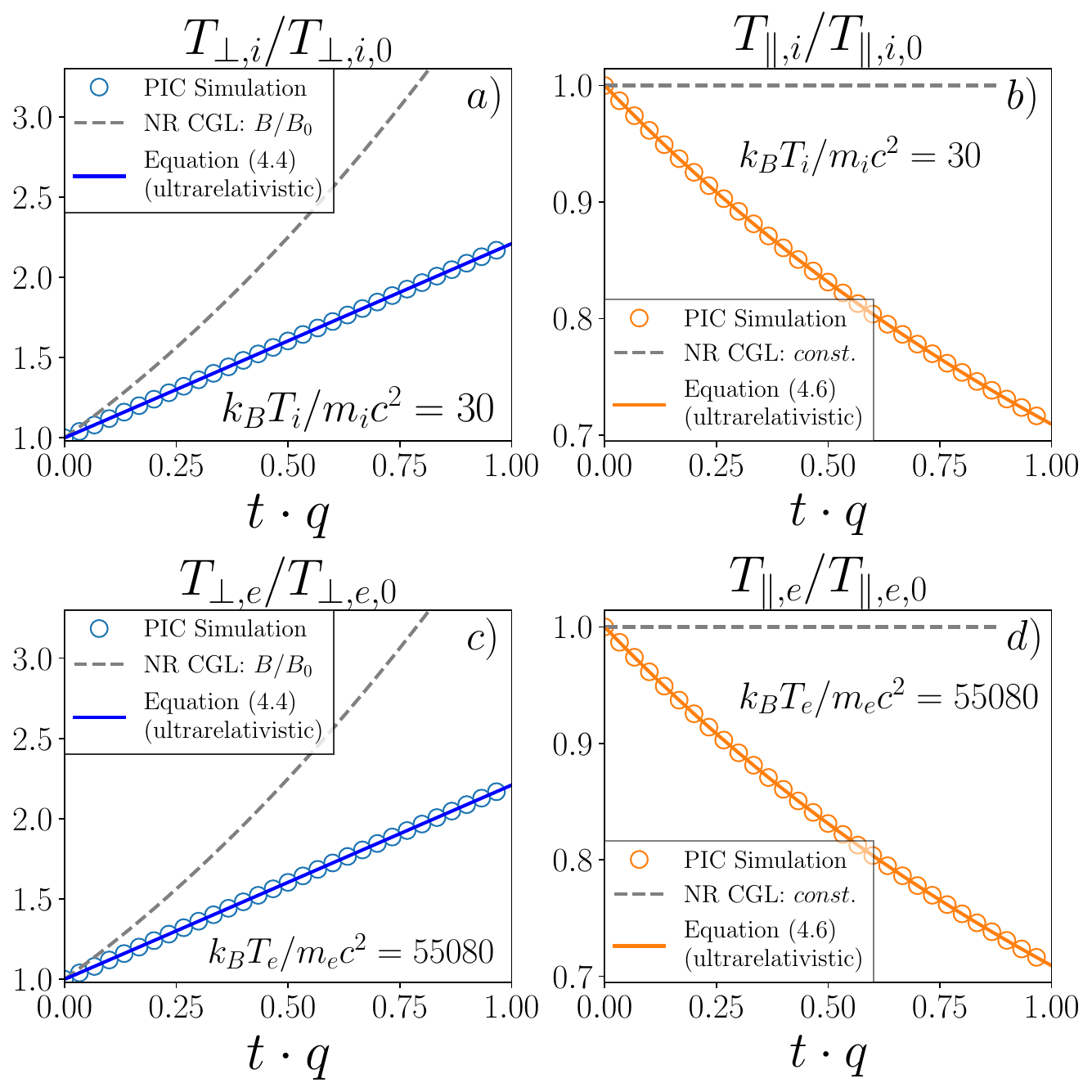}
    \caption{Panel $a$: The evolution of the ion perpendicular temperature for run Compb0.5m1836d30wcis3200 (open blue circles), for initial $k_BT_i^{\text{init}}/m_ic^2=30$. The solid blue line shows the evolution of $T_{\perp}$ according to eqn. \eqref{eq:Pperp_compressing_UR}. Panel $b$: The evolution of the ion parallel temperature (open orange circles). The solid orange line shows the evolution of $T_{\parallel}$ according to eqn. \eqref{eq:Ppar_compressing_UR}. In panels $a$ and $c$ the nonrelativistic CGL evolution for $P_{\perp}$ is shown in dashed gray line for a compressing motion. In panels $b$ and $d$, the nonrelativistic CGL evolution for $P_{\parallel}$ is shown in dashed gray line for a compressing motion.}
    \label{fig:Comp_delgam30_ion}
\end{figure}

The evolution of the ion perpendicular temperature $T_{\perp,i}$ and parallel temperature $T_{\parallel,i}$, and the electron perpendicular and parallel temperatures $T_{\perp,e}$, $T_{\parallel,e}$ are shown in figure \ref{fig:Comp_delgam30_ion} for run Compb0.5m8d0.2wcis800. Here we can test our ultrarelativistic double-adiabatic equations \eqref{eq:Pperp_compressing_UR} and \eqref{eq:Ppar_compressing_UR}.

Similar to the shearing case, we can see that for this ultrarelativistic initial temperature both $T_{\perp,i}$ (open blue circles in panel $a)$) and $T_{\parallel,i}$ (open orange circles in panel $b$) also depart from their nonrelativistic CGL prediction (dashed gray lines), especially for $T_{\perp,i}$, which shows a more pronounced deviation than in the mildly relativistic case, due to the high initial temperature.  

Nevertheless, we also see here that our ultrarelativistic double-adiabatic equations \eqref{eq:Pperp_compressing_UR} (solid blue line in panel $a$) and \eqref{eq:Ppar_compressing_UR} (solid orange line in panel $b$) consistently follow the evolution of $T_{\perp,i}$ and $T_{\parallel,i}$ from the simulation.

The evolution of the electron perpendicular temperature $T_{\perp,e}$ and parallel temperature $T_{\parallel,e}$ for run Compb0.5m8d0.2wcis800 is shown in figure \ref{fig:Comp_delgam30_ion}. 

For electrons in run Compb0.5m1836d30wcis3200, the behavior is analogous to the corresponding shearing simulation. We see that the departure from nonrelativistic CGL predictions (dashed gray lines) is present in both $P_{\perp,e}$ (open blue circles, panel $c$) and $P_{\parallel,e}$ (open orange circles, panel $d$). For this very high $k_BT_e^{\text{init}}/m_ec^2$, we can see that our ultrarelativistic double-adiabatic equations \eqref{eq:Pperp_compressing_UR} (solid blue line, panel $c$), \eqref{eq:Ppar_compressing_UR} (solid orange line, panel $d$) also work very well in describing the evolution of $P_{\perp,e}$ and $P_{\parallel,e}$ in the simulations.

\subsection{Distribution function comparison}

From both shearing and compressing simulations, we can obtain the ion and electron distribution functions at any time by constructing the histograms of the particle momenta $p_{\perp,\alpha}$, $p_{\parallel,\alpha}$ ($\alpha=i,e$). We can then compare these empirical distributions with our solution \eqref{eq:SolutionDistributionFunction_MJ_Cyl} of the drift kinetic equation \eqref{eq:DriftKineticEquationCyl}, to assess how well it describes the evolution of the distribution in the simulations, from an initial, isotropic Maxwell-Jüttner into its anisotropic form. 

\begin{figure}
    \centering
    \begin{tabular}{cc}
         \includegraphics[width=0.95\linewidth]{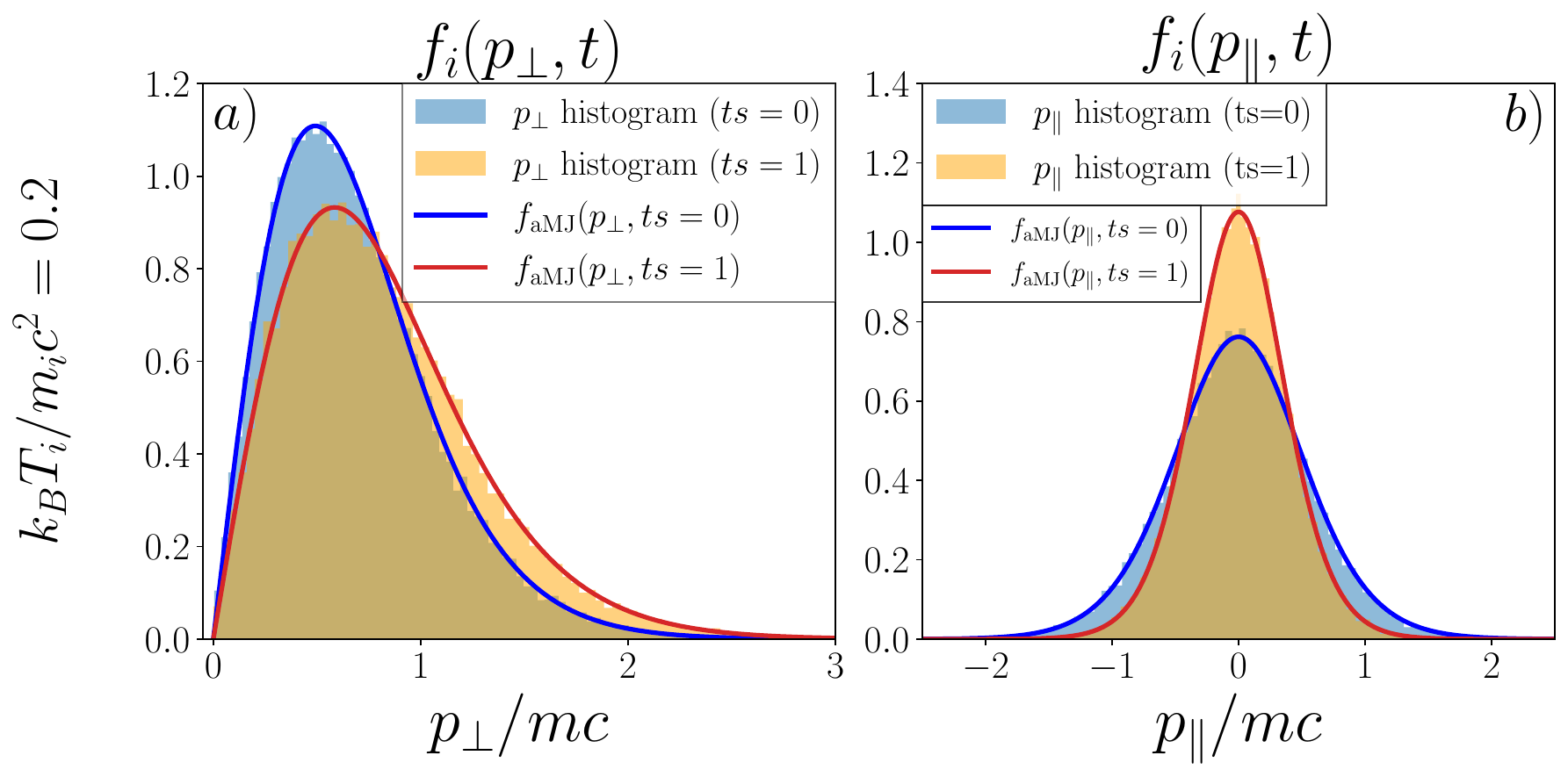}  \\
         \includegraphics[width=0.95\linewidth]{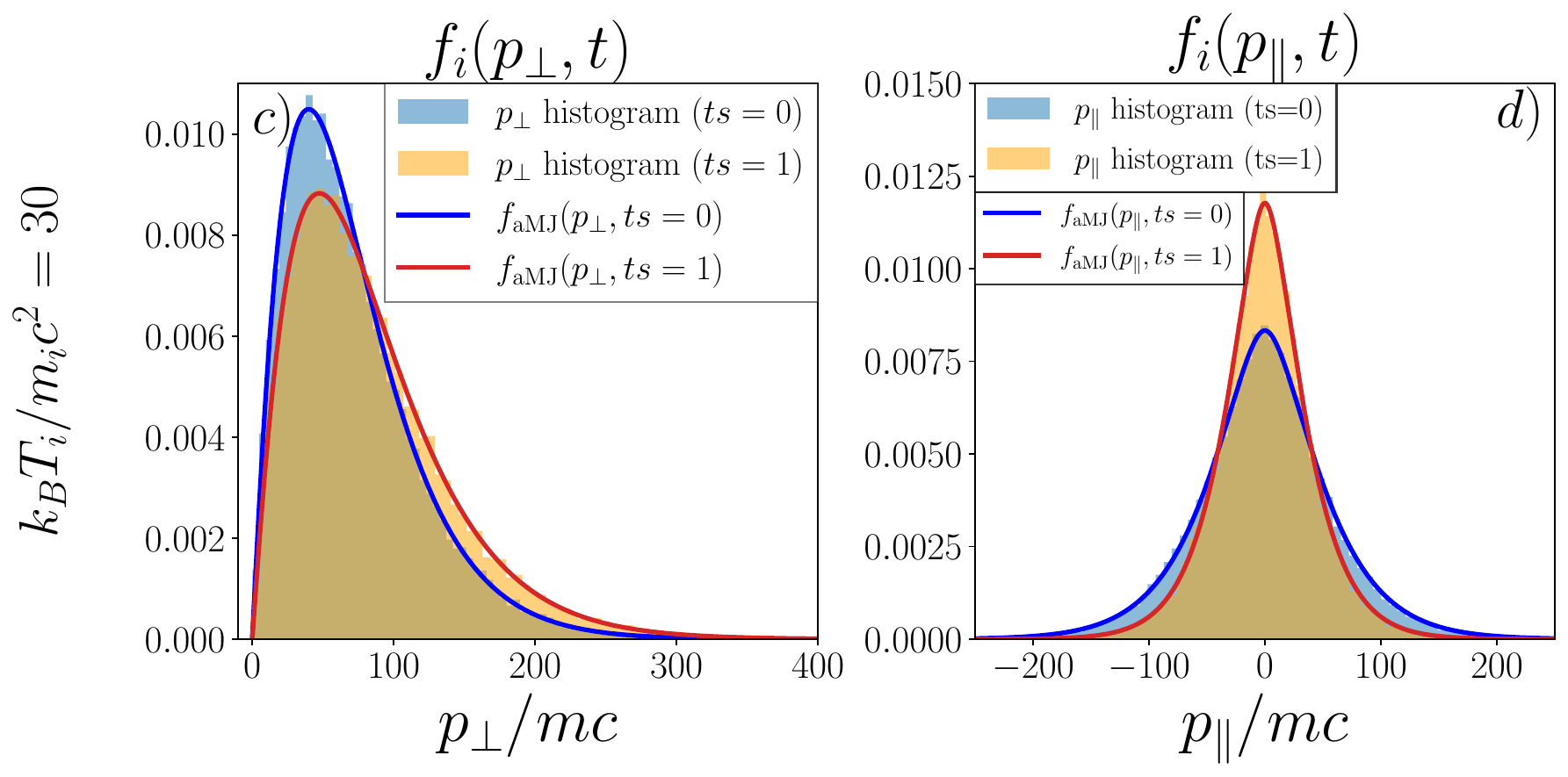} 
    \end{tabular}
    \caption{Panel $a$: The histograms of the ion perpendicular momentum for run Shb0.5m8d0.2wcis800 at $t\cdot s = 0$ (light blue bars) and $t\cdot s = 1$ (yellow bars). The marginal distribution $f_{\text{aMJ}}(p_{\perp},t)$, eqn. \eqref{eq:fperp_marginal}, is shown at $t\cdot s = 0$ (solid blue line) and $t\cdot s = 1$ (solid red line), for $\theta=0.2$. Panel $b$: The histograms of the ion parallel momentum at $t\cdot s = 0$ (light blue bars) and $t\cdot s = 1$ (yellow bars). The marginal distribution $f_{\text{aMJ}}(p_{\parallel},t)$, eqn. \eqref{eq:fpar_marginal}, is shown at $t\cdot s = 0$ (solid blue line) and $t\cdot s = 1$ (solid red line), for $\theta=0.2$. Panels $c$ and $d$ show the same quantities as panels $a$ and $b$, respectively, but for run Shb0.5m8d30wcis3200 (i.e. $\theta=30$).}
    \label{fig:Distribution_Comparison_ion}
\end{figure}

From our time-dependent distribution function, equation \eqref{eq:SolutionDistributionFunction_MJ_Cyl}, 

\begin{align}
    \nonumber
    f(t,p_{\perp},p_{\parallel}) = \frac{n_0}{4\pi m^3c^3 \theta K_2(1/\theta)}\exp\left(-\frac{1}{\theta}\sqrt{1 + \frac{p_{\perp}^2/m^2c^2}{(B/B_0)} + \frac{p_{\parallel}^2/m^2c^2}{(n/n_0)^{2}(B/B_0)^{-2}}}\right),
\end{align}

we can obtain the marginal distributions in $p_{\perp}$ and $p_{\parallel}$ by integrating,

\small
\begin{align}
    f_{\text{aMJ}}(t,p_{\perp}) &= \frac{n_0}{4\pi m^3c^3 \theta K_2(1/\theta)} \int_{-\infty}^{\infty} \exp\left(-\frac{1}{\theta}\sqrt{1 + \frac{p_{\perp}^2/m^2c^2}{(B/B_0)} + \frac{p_{\parallel}^2/m^2c^2}{(n/n_0)^{2}(B/B_0)^{-2}}}\right) 2\pi p_{\perp} dp_{\parallel}
\end{align}
\normalsize
\begin{align}
    f_{\text{aMJ}}(t,p_{\perp}) &=  \frac{n_0}{m^2c^2 \theta K_2(1/\theta)} \frac{(n/n_0)}{(B/B_0)}p_{\perp}\sqrt{1 + \frac{p_{\perp}^2/m^2c^2}{(B/B_0)}} K_1\left(\frac{1}{\theta}\sqrt{1 + \frac{p_{\perp}^2/m^2c^2}{(B/B_0)}}\right). 
    \label{eq:fperp_marginal}
\end{align}

and

\small
\begin{align}
    f_{\text{aMJ}}(t,p_{\parallel}) &= \frac{n_0}{4\pi m^3c^3 \theta K_2(1/\theta)} \int_{0}^{\infty} \exp\left(-\frac{1}{\theta}\sqrt{1 + \frac{p_{\perp}^2/m^2c^2}{(B/B_0)} + \frac{p_{\parallel}^2/m^2c^2}{(n/n_0)^{2}(B/B_0)^{-2}}}\right) 2\pi p_{\perp} dp_{\perp}\\
    f_{\text{aMJ}}(t,p_{\parallel}) &= \frac{n_0}{2mc K_2(1/\theta)}\left(\sqrt{1 + \frac{p_{\parallel}^2/m^2c^2}{(n/n_0)^2(B/B_0)^{-2}}} + \theta \right)\exp\left( -\frac{1}{\theta}\sqrt{1 + \frac{p_{\parallel}^2/m^2c^2}{(n/n_0)^2(B/B_0)^{-2}}} \right)
    \label{eq:fpar_marginal}
\end{align}
\normalsize

where $K_1(x)$ is the modified Bessel function of the first kind. The comparison between the histograms of ion perpendicular momentum $p_{\perp,i}$ with the marginal distribution $f_{\text{aMJ}}(t,p_{\perp})$ and the ion parallel momentum $p_{\parallel,i}$ with $f_{\text{aMJ}}(t,p_{\parallel})$ for our shearing simulations is shown in figure \ref{fig:Distribution_Comparison_ion} for initial $k_BT_i^{\text{init}}/m_ic^2 = 0.2$ and $k_BT_i^{\text{init}}/m_ic^2 = 30$. This is shown at the beginning and at the end of simulations (i.e. from initial isotropic state at $t\cdot s=0$ to an anisotropic state at $t\cdot s= 1$). We are showing ion results only, as electron results show a very similar behavior.

We can see that for all cases presented, the agreement between simulations and theory is noteworthy. For the perpendicular momentum (panels $a$ and $c$ in fig. \ref{fig:Distribution_Comparison_ion}) we can see very good agreement at $t\cdot s = 0$ between the $p_{\perp,i}$ histogram (light blue bars) and our analytical marginal distribution $f_{\text{aMJ}}(ts=0,p_{\perp})$ (eqn. \eqref{eq:fperp_marginal}, solid blue line), validating the isotropic limit to the Maxwell-Jüttner distribution. More notably, at the end of the simulation we see the expected widening of the $p_{\perp,i}$ histogram (yellow bars) towards higher values, transitioning to an anisotropic state, and our marginal distribution $f_{\text{aMJ}}(ts=1,p_{\perp})$ (solid red line) shows very good agreement with the simulation at this stage. 

For the parallel momentum (panels $b$ and $d$ in fig. \ref{fig:Distribution_Comparison_ion}) we also obtain very good correspondence at $t\cdot s = 0$ between $p_{\parallel,i}$ (light blue bars) and the analytical marginal distribution $f_{\text{aMJ}}(ts=0,p_{\parallel})$ (eqn. \eqref{eq:fpar_marginal}, solid blue line), and we validate the isotropic limit in this case as well. At $t\cdot s = 1$, we see the expected narrowing of the parallel momentum distribution (yellow bars) towards smaller values, and in this case we see that our marginal distribution $f_{\text{aMJ}}(ts=1,p_{\parallel})$ (solid red line) correctly characterizes this anisotropic state as well.

\begin{figure}
    \centering
    \begin{tabular}{cc}
        \includegraphics[width=0.46\linewidth]{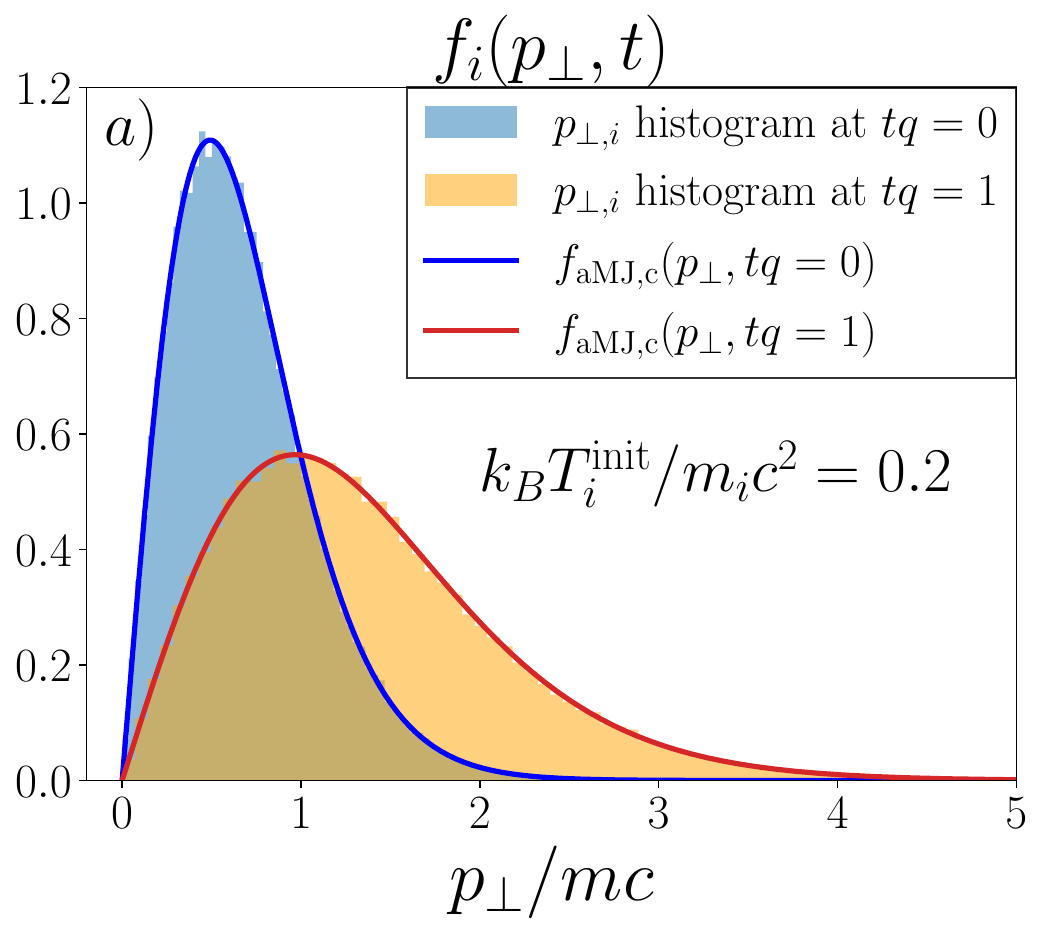}
         & 
         \includegraphics[width=0.48\linewidth]{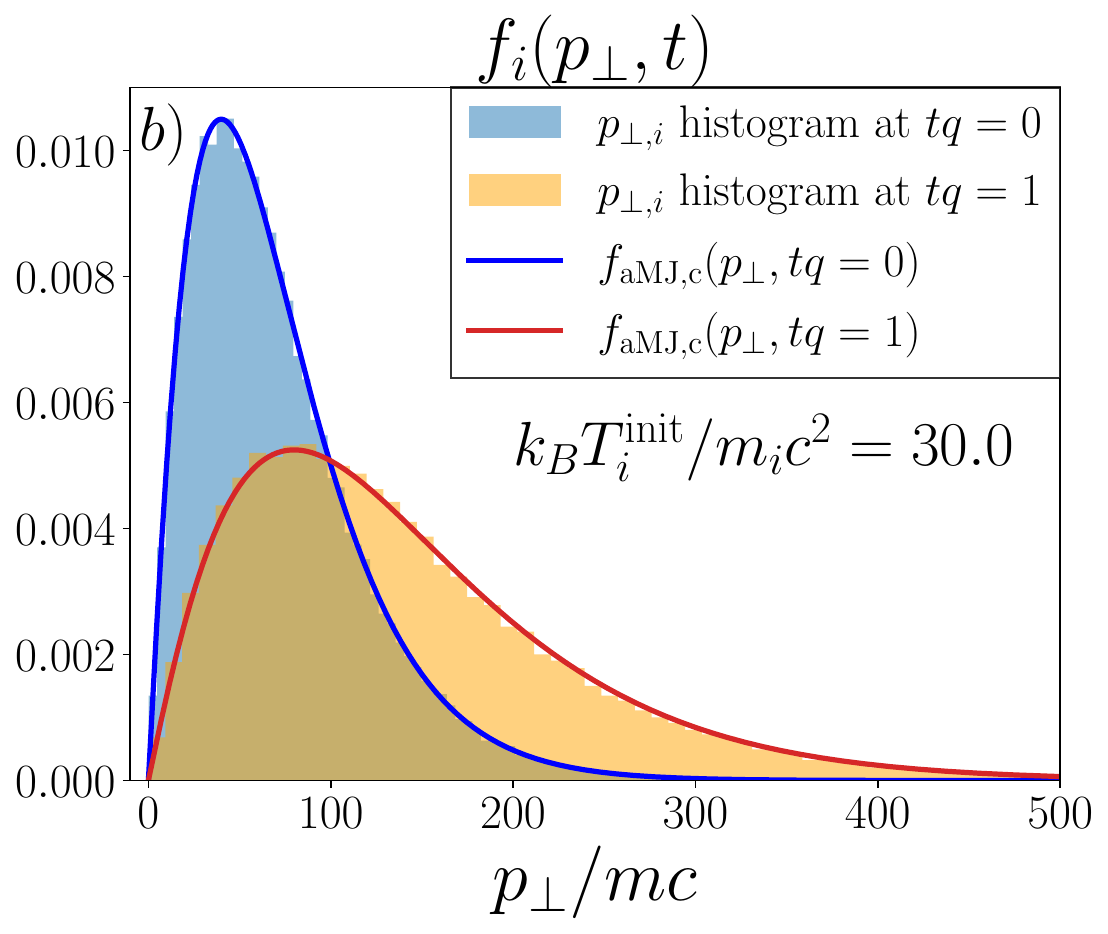}
    \end{tabular}
    \caption{Panel $a$: The histograms of the ion perpendicular momentum for run Compb0.5m8d0.2wcis800 at $t\cdot s = 0$ (light blue bars) and $t\cdot s = 1$ (yellow bars). The marginal distribution $f_{\text{aMJ,c}}(p_{\perp},t)$, eqn. \eqref{eq:fperp_marginal_compress}, is shown at $t\cdot s = 0$ (solid blue line) and $t\cdot s = 1$ (solid red line), for $\theta=0.2$. Panel $b$: The histograms of the ion perpendicular momentum for run Compb0.5m8d30wcis3200 at $t\cdot s = 0$ (light blue bars) and $t\cdot s = 1$ (yellow bars). The marginal distribution $f_{\text{aMJ,c}}(p_{\perp},t)$, eqn. \eqref{eq:fperp_marginal_compress}, is shown at $t\cdot s = 0$ (solid blue line) and $t\cdot s = 1$ (solid red line), for $\theta=30$.}
    \label{fig:Distribution_Comparison_ion_compress}
\end{figure}

For our compressing simulations, recall the solution \eqref{eq:Solution_DKE_Compressing_MJ}

\begin{align}
    f(t,p_{\perp},p_{\parallel}) = \frac{n_0}{4\pi m^3c^3 \theta K_2(1/\theta)}\exp\left( -\frac{1}{\theta}\sqrt{1 + \frac{p_{\perp}^2/m^2c^2}{(B/B_0)} + p_{\parallel}^2/m^2c^2} \right). 
\end{align}

The marginal distribution $f_{\text{aMJ,c}}(t,p_{\perp})$ can be calculated as 

\begin{align}
    f_{\text{aMJ,c}}(t,p_{\perp}) &= \frac{n_0}{4\pi m^3c^3 \theta K_2(1/\theta)} \int_{-\infty}^{\infty} \exp\left(-\frac{1}{\theta}\sqrt{1 + \frac{p_{\perp}^2/m^2c^2}{(B/B_0)} + p_{\parallel}^2/m^2c^2}\right) 2\pi p_{\perp} dp_{\parallel}\\
    f_{\text{aMJ,c}}(t,p_{\perp}) &= \frac{n_0}{m^2c^2K_2(1/\theta)}p_{\perp}\sqrt{1 + \frac{p_{\perp}^2/m^2c^2}{B/B_0}}K_1\left(\frac{1}{\theta}\sqrt{1 + \frac{p_{\perp}^2/m^2c^2}{B/B_0}}\right)
    \label{eq:fperp_marginal_compress}
\end{align}

The comparison between the histograms of ion perpendicular momentum $p_{\perp,i}$ with $f_{\text{aMJ,c}}(t,p_{\perp})$ for the compressing simulations is shown in figure \ref{fig:Distribution_Comparison_ion_compress}, also for $k_BT_i^{\text{init}}/m_ic^2 = 0.2$ and $k_BT_i^{\text{init}}/m_ic^2 = 30$, and at $t\cdot q = 0$ and $t\cdot q = 1$. We only show the perpendicular momentum, as we know that  the parallel momentum of particles does not evolve. Electron results are not shown but are completely analogous.

The agreement between $f_{\text{aMJ,c}}(t,p_{\perp})$ and $p_{\perp,i}$ from simulations is also remarkable in this case, for both initial temperature cases. At $t\cdot q = 0$, the histogram of the perpendicular momentum (light blue bars) is very well described by $f_{\text{aMJ,c}}(tq=0,p_{\perp})$ (solid blue line), validating the isotropic limit of our solution in this case as well. At the end of the simulation, we can clearly see the expected spread of the $p_{\perp,i}$ distribution (yellow bars) towards higher values, and the marginal $f_{\text{aMJ,c}}(tq=1,p_{\perp})$ (solid red line) describes this anisotropic state very accurately as well.

We have therefore demonstrated that the distribution function \eqref{eq:SolutionDistributionFunction_MJ}, \eqref{eq:SolutionDistributionFunction_MJ_Cyl}, which is a solution to the drift kinetic equation \eqref{eq:DriftKineticEquationCyl}, correctly describes the double-adiabatic evolution of the distribution in shearing and compressing PIC simulations at any time. Therefore, these distributions constitute a newly proposed form for a time-dependent, anisotropic Maxwell-Jüttner distribution, derived from first principles.

\section{Discussion}
\label{sec:Discussion}

The results reported in this work are relevant for a diversity of high-energy astrophysical phenomena involving relativistically hot, collisionless plasmas, both low and high beta. Among these systems one can mention pulsar wind nebulae, low-luminosity accretion flows around SMBH, relativistic jets from SMBH in active galaxies like blazars, and also cosmic rays. In absence of heat fluxes and where spatial gradients are much larger than the gyroradii of particles, the relativistic species present in these plasmas can be modeled by the distribution functions \eqref{eq:SolutionDistributionFunction_MJ}, \eqref{eq:SolutionDistributionFunction_MJ_Cyl}, and \eqref{eq:SolutionDistributionFunctionUR} for ultrarelativistic temperatures. Additionally, the adiabatic evolution of the perpendicular and parallel pressures associated to these relativistic species are well described by equations \eqref{eq:Pperp_MJ}, \eqref{eq:Ppar_MJ}, and \eqref{eq:Pperp_UR},\eqref{eq:Ppar_UR}. This evolution will be valid until kinetic microinstabilities are excited and start to provide an effective collisionality in the plasma via complex wave-particle interactions.

\subsection{Relativistic Extended MHD in the Braginskii Limit}
A direct application of these results is in the context of extended magnetohydrodynamic theories for relativistic collisionless plasmas (e.g. \citet{TenBarge2008,Chandra2015}). The distribution function presented in this work, along with the relativistic double-adiabatic equations, can be used as a closure to the MHD fluid system while retaining pressure anisotropies in the stress-energy tensor, provided that the system has no heat fluxes and spatial gradients in $B$ and $n$ are much larger than the particles' gyration radius.

Interestingly, we can extend the results in \citet{Chandra2015} for the evolution of the pressure anisotropy $\Delta P = P_{\perp}-P_{\parallel}$ in the ultrarelativistic case, and apply it to the collisional Braginskii limit \citep{Braginskii1965}. Considering the $R>1$ case only, from equations \eqref{eq:Pperp_UR},\eqref{eq:Ppar_UR}, and $P=(2P_{\perp} + P_{\parallel})/3$ we can write

\begin{align}
    \frac{1}{P}\frac{d\Delta P}{dt} &=  \mathcal{A}\frac{\dot{B}}{B} + \mathcal{C}\frac{\dot{n}}{n},
    \label{eq:AnisotropyEvolution}
\end{align}

where 

\begin{align}
    \mathcal{A} &= \frac{3}{2}\Psi(r)\left[ \frac{(r^3 + 2)(r^3 + 8)}{2(r^3 - 1)^{5/2}}\arctan\left( \sqrt{r^3 - 1} \right) + \frac{r^6 - 32r^3 + 4}{2r^3(r^3-1)^2} \right],\\
    \mathcal{C} &= -3\Psi(r)\left[ \frac{r^3(10-r^3)}{2(r^3 - 1)^{5/2}}\arctan(\sqrt{r^3 - 1}) - \frac{r^6 + 12r^3 - 4}{2r^3(r^3-1)^2} \right],\\
    \Psi(r) &= \frac{r^3\sqrt{r^3 - 1}}{r^3\arctan(\sqrt{r^3 - 1}) + \sqrt{r^3 - 1}},
\end{align}

and $r \equiv R/R_0$. Equation \eqref{eq:AnisotropyEvolution} is valid for any value of anisotropy (as long as the evolution is adiabatic, see below). Note that when we go to the small anisotropy limit, corresponding to $r \rightarrow 1$ (i.e. $n$ and $B$ close to their initial values), we exactly recover the result in \citet{Chandra2015}. Indeed, as $\Psi(r\rightarrow 1)\rightarrow 1/2$, $\mathcal{A} \rightarrow 12/5$, and $\mathcal{C}\rightarrow 8/5$, and we get

\begin{align}
    \frac{1}{P}\frac{d\Delta P}{dt} =  \frac{12}{5}\frac{\dot{B}}{B} - \frac{8}{5}\frac{\dot{n}}{n} = \frac{4}{5}\frac{d}{dt}\ln\left(\frac{B^3}{n^2}\right) + \mathcal{O}\left([\Delta P/P]^2\right).
\end{align}

In the Braginskii limit, the adiabatic evolution of the pressure anisotropy is balanced with the scattering process present in the plasma. Denoting the scattering rate as $\nu$ in the collisional limit, we can write, in equilibrium,

\begin{align}
    \Delta P = \frac{P}{\nu}\left[ \mathcal{A}\frac{\dot{B}}{B} - \mathcal{C}\frac{\dot{n}}{n} \right].
    \label{eq:AnisotropyBraginskii}
\end{align}

Equation \eqref{eq:AnisotropyBraginskii} is then valid for any values of $B(t)$ and $n(t)$, in the ultrarelativistic regime ($k_BT/mc^2 \gtrsim 1$, in practice).

\subsection{Small Amplitude Limit }
 
 The analysis up to now is valid for arbitrarily large changes in $B$ and $n$, provided the changes are slow and the plasma is collisionless, due to the isotropizing effect of collisions. However, even when binary collisions are negligible, particles can be scattered by interactions with waves, and a case of particular interest is scattering by waves generated by the particles themselves, through kinetic scale anisotropy instabilities. When the threshold for these instabilities is low, the evolution up to the triggering of instability can be calculated from perturbation theory. For example,  cosmic rays trigger a gyroresonant instability of Alfven waves when the fractional pressure anisotropy is of order $v_A/c$ \citep[e.g.][]{Zweibel2020}.
 
Suppose we replace $R/R_0$ and $n/n_0$ by $1+\delta R/R$ and $1+\delta n/(3n)$, respectively, where $\vert\delta R/R\vert$ and $\vert\delta n/n\vert$ are $\ll 1$. And, assume that $f_0$ is isotropic. Then eqn. (\ref{eq:SolutionDistributionFunctionSph}) can be replaced by a Taylor expansion, which to first order in the perturbation amplitude is
\begin{equation}\label{eq:taylor1}
f(p,\mu)=f_0(p)+\left( \frac{\delta R}{R}P_2(\mu)-\frac{\delta n}{3n}P_0(\mu) \right)p\frac{df_0}{dp},
\end{equation}
where $P_2$ and $P_0$ are Legendre functions of order 2 and 0, respectively. 

Equation (\ref{eq:taylor1}) is in convenient form for calculation of the pressure anisotropy,
\begin{equation}\label{eq:DeltaP1}
\Delta P\equiv P_{\perp} - P_{\parallel}=-\int p^2dp d\mu pvP_2(\mu)f(p,\mu).
\end{equation}
Using eqn. (\ref{eq:taylor1}) in eqn. (\ref{eq:DeltaP1}) we have
\begin{equation}\label{eq:DeltaP2}
\Delta P =-\frac{2}{5}\frac{\delta R}{R}\int p^4v\frac{df_0}{dp}.
\end{equation}
For the important example of cosmic rays, we take the ultrarelativistic limit $v\rightarrow c$. Then, integrating eqn. (\ref{eq:DeltaP2}) by parts and using the standard definition of pressure, we find
\begin{equation}\label{eq:DeltaP3}
\Delta P =\frac{12}{5}\frac{\delta R}{R}P.
\end{equation}

\section{Conclusions}
\label{sec:conclusions}
In this work, we have extended the double adiabatic equations for a homogeneous, collisionless plasma to the relativistic regime, and presented a novel anisotropic version of the Maxwell-Jüttner distribution. We analytically solved the drift kinetic equation in the homogeneous, collisionless limit, and obtained a general, time-dependent solution for the distribution function in terms of density and magnetic field variations.

By considering a relativistic Maxwell-Jüttner distribution as initial condition, we obtained new, relativistic expressions for the evolution of both perpendicular and parallel pressures, in terms of the integral moments that can be solved numerically for any temperature $\theta = k_BT/mc^2$. When considering an ultrarelativistic Maxwell-Jüttner distribution as initial condition, the integral moments are tractable, and we obtained analytical evolution equations for the perpendicular and parallel pressures. In practice, these analytical expressions prove to be applicable for $k_BT/mc^2 \geq 1$. We confirmed that our relativisic double-adiabatic equations agree with the general state equations for a relativistic anisotropic plasma from \citet{Gedalin1991}, and we validated them using PIC simulations of shearing and compressing boxes for mildly relativistic and ultrarelativistic initial temperatures, obtaining remarkable agreement. From a general perspective, we conclude that the specific form of the double-adiabatic equations then depends on the initial distribution function that is used.

Additionally, the time-dependent solution eqn. \eqref{eq:SolutionDistributionFunction_MJ_Cyl} obtained for an initial Maxwell-Jüttner distribution constitutes a new anisotropic extension for a relativistic Maxwellian, explicitly dependent on the density and magnetic field variations. For a specific value of the pressure anisotropy, the corresponding double-adiabatic equations can be used to map the density and magnetic field variations to the specific level of anisotropy. We confirmed that this distribution correctly describes the momentum distributions of both shearing and compressing PIC simulations, in both initial isotropic and final anisotropic states. As PIC simulations utilize a discretized version of the method of characteristics in their algorithms, we argue that our distribution \eqref{eq:SolutionDistributionFunction_MJ} is the analytical form of the distribution function that our shearing and compressing box simulations evolve (when the initial condition is a Maxwell-Jüttner distribution), until an instability is excited. This analytical anisotropic Maxwell-Jüttner form could also be used to qualitatively assess the distortions to the numerical distribution function as a consequence of wave-particle interactions by microinstabilities in driven PIC simulations.

These new findings open several paths for future work. The relativistic double-adiabatic equations can be used as a new closure for extended models of relativistic magnetohydrodynamic (MHD) equations (\citet{TenBarge2008,Chandra2015}) for systems where heat fluxes are negligible and are well magnetized so the particle gyroradius is much smaller than the relevant spatial gradients present in the plasma. This provides an interesting framework that can be relevant for general relativistic MHD simulations used in astrophysical  studies of, e.g., low-luminosity accretion disks around supermassive black holes \citep{Akiyama2019}, pressure anisotropy effects on cosmic ray hydrodynamics \citep{Zweibel2020}, and for extending CGL-MHD fluid models like those used for studying magneto-immutability effects \citep{Majeski2024} in the intracluster medium (ICM) of galaxy clusters, in order to include relativistic species that are part of the ICM (e.g. cosmic rays). Additionally, the linear stability of our time-dependent anisotropic Maxwell-Jüttner can then be studied analytically and numerically (e.g. using linear solvers like ALPS \citep{Verscharen2018}), along with the onset and evolution of pressure anisotropy driven microinstabilities relevant in the relativistic regime. The results can then be compared directly with driven PIC simulations. A direct implementation of this anisotropic Maxwell-Jüttner distribution in PIC simulations is also possible, and will be addressed in a future work. 

Finally, our general solution to the drift kinetic equation also admits other types of interesting initial distributions, including power-law and kappa distributions, relevant for cosmic-ray and solar wind studies, respectively. Double-adiabatic equations and linear stability analysis can also be performed in each case.

\section*{Acknowledgments}

    The authors thank the anonymous referees for the insightful comments and suggestions that improved the paper. We thank Mario Riquelme, Astor Sandoval, and Martín Astete for useful comments and discussions. During the final stages of preparation of this manuscript, we became aware of a similar and independent study by A. Wierzchucka, P. J. Bilbao, A. G. R. Thomas, D. A. Uzdensky, and A. A. Schekochihin (accepted for publication in JPP), carried out independently and simultaneously to ours \citep{Wierzchucka2026}. The authors thank Vedant Dhruv for pointing out a typo in the exponent in one of the equations, that was corrected in this version of the paper.

\section*{Funding}

F.L acknowledges support by the International Space Science Institute (ISSI) in Bern, through ISSI International Team project 24‐612 (“Excitation and dissipation of kinetic‐scale fluctuations in space plasmas”).
A.T. was supported by the Department of Energy Fusion Energy Sciences Postdoctoral Research Program, administered by the Oak Ridge Institute for Science and Education and Oak Ridge Associated Universities under DOE contract DE-SC0014664.
E.G.Z. acknowledges support from the National Science Foundation through Grant PHY-2409224 and the Simons Foundation through the Simons Collaboration on Extreme Electrodynamics of Compact Sources as well as the hospitality of the Leonard Parker Center for  Gravitation, Cosmology, and Astrophysics at UW--Milwaukee, where some of this work was completed.

\section*{Declaration of interests}

The authors report no conflict of interest.

\appendix
\section{}\label{appA}

\subsection{$m_i/m_e$ and $\beta_i^{\text{init}}$ dependency in Shearing Simulations}

\begin{figure}
    \centering
    \includegraphics[width=0.95\linewidth]{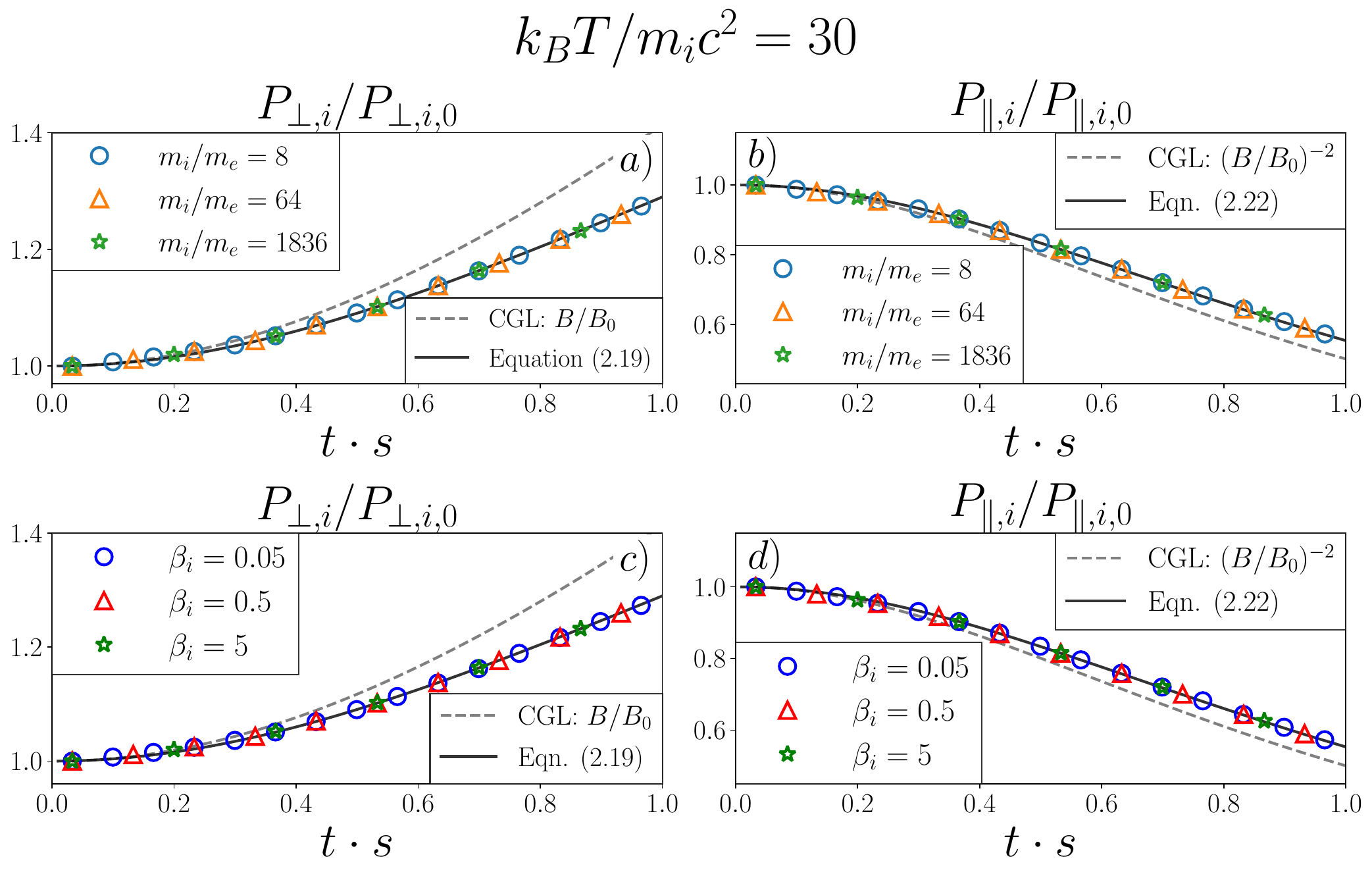}
    \caption{Panel $a$: The evolution of the ion perpendicular pressure for runs Shb0.5m8d30wci3200 (pale blue circles), Shb0.5m64d30wci3200 (orange triangles), and Shb0.5m1836d30wci3200 (green stars). Panel $b$: The evolution of the ion parallel pressure for runs Shb0.5m8d30wci3200 (pale blue circles), Shb0.5m64d30wci3200 (orange triangles), and Shb0.5m1836d30wci3200 (green stars). Panel $c$: The evolution of the ion perpendicular pressure for runs Shb0.05m8d30wci3200 (blue circles), Shb0.5m8d30wci3200 (red triangles), and Shb5m8d30wci3200 (green stars). Panel $d$: The evolution of the ion parallel pressure for runs Shb0.05m8d30wci3200 (blue circles), Shb0.5m8d30wci3200 (red triangles), and Shb5m8d30wci3200 (green stars). In panels $a$ and $c$, the nonrelativistic CGL evolution for $P_{\perp}$ is shown in dashed gray line, and the ultrarelativistic evolution for $P_{\perp}$ (cf. eqn. \eqref{eq:Pperp_UR}) is shown in solid black line. In panels $b$ and $d$, the nonrelativistic CGL evolution for $P_{\parallel}$ is shown in dashed gray line, and the ultrarelativistic evolution for $P_{\parallel}$ (cf. eqn. \eqref{eq:Ppar_UR}) is shown in solid black line.}
    \label{fig:beta_mime_comparison}
\end{figure}

In our shearing simulations, we tested various cases for both the ion to electron mass ratio $m_i/m_e$ and initial plasma beta $\beta_i^{\text{init}}$. We saw no dependency on any of these two parameters, analogous to the nonrelativistic CGL case. These results are collected in figure \ref{fig:beta_mime_comparison}. We are showing ion quantities only, but the evolution of electron pressures is analogous.

For the mass ratio, we tried three different values: $m_i/m_e = 8, 64, 1836$ (panels $a$ and $b$ in fig. \ref{fig:beta_mime_comparison}). For both ion perpendicular and parallel pressures, we can see that all three mass ratio cases are well described by their respective ultrarelativistic evolution (cf. eqn. \eqref{eq:Pperp_UR},\eqref{eq:Ppar_UR}).

Similarly, for the initial beta, we tried three different values spanning two orders of magnitude: $\beta_i^{\text{init}}= 0.05, 0.5,$ and $ 5$ (panels $c$ and $d$ in fig. \ref{fig:beta_mime_comparison}). For both ion perpendicular and parallel pressures, their evolution is well explained by their respective ultrarelativistic evolution in all three cases.

\subsection{$m_i/m_e$ and $\beta_i^{\text{init}}$ dependency in Compressing Simulations}

\begin{figure}
    \centering
    \includegraphics[width=0.95\linewidth]{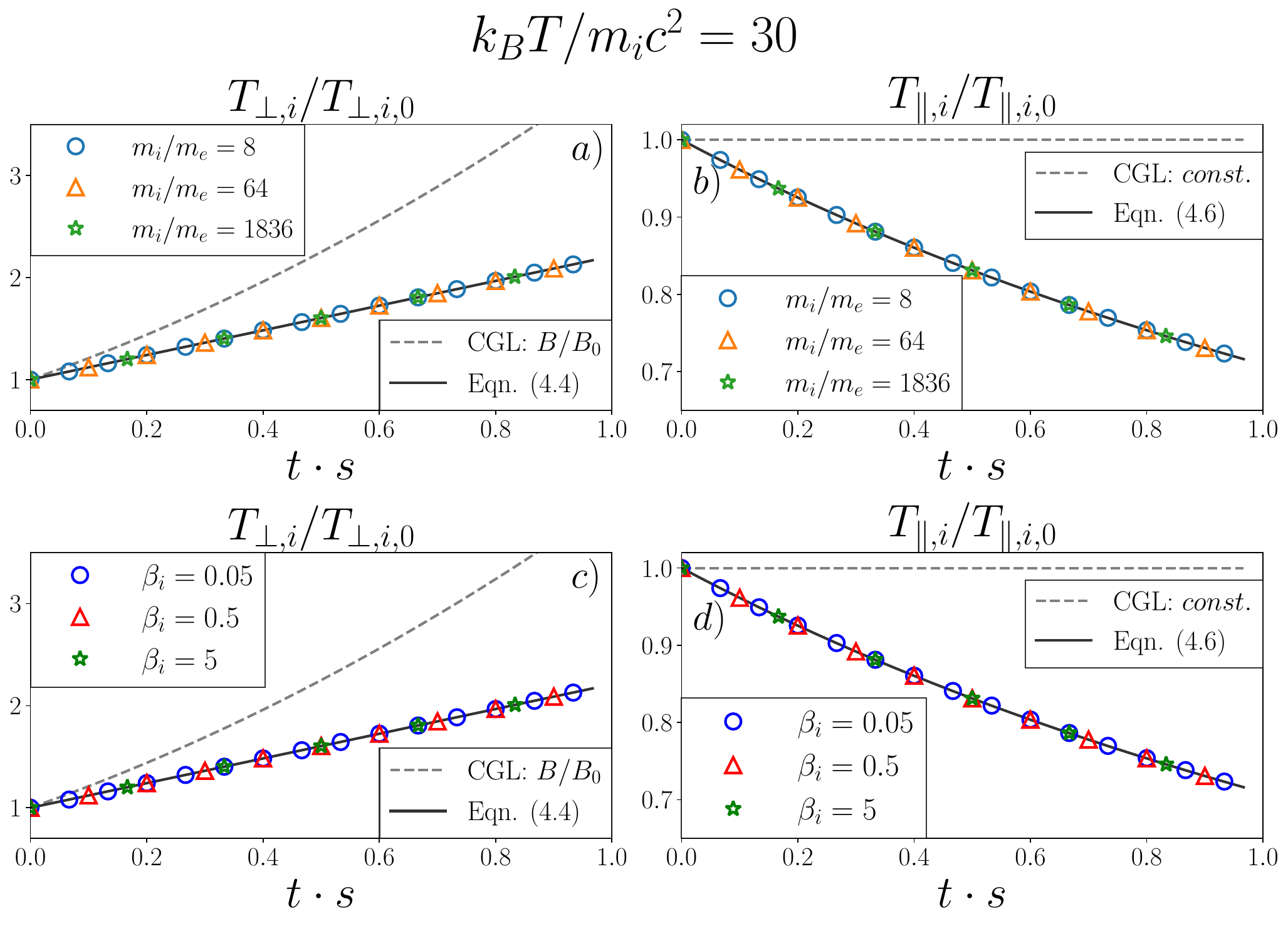}
    \caption{Panel $a$: The evolution of the ion perpendicular temperature for runs Compb0.5m8d30wci3200 (pale blue circles), Compb0.5m64d30wci3200 (orange triangles), and Compb0.5m1836d30wci3200 (green stars). Panel $b$: The evolution of the ion parallel temperature for runs Compb0.5m8d30wci3200 (pale blue circles), Compb0.5m64d30wci3200 (orange triangles), and Compb0.5m1836d30wci3200 (green stars). Panel $c$: The evolution of the ion perpendicular temperature for runs Compb0.05m8d30wci3200 (blue circles), Compb0.5m8d30wci3200 (red triangles), and Compb5m8d30wci3200 (green stars). Panel $d$: The evolution of the ion parallel temperature for runs Compb0.05m8d30wci3200 (blue circles), Compb0.5m8d30wci3200 (red triangles), and Compb5m8d30wci3200 (green stars). In panel $a$ and $c$, the nonrelativistic CGL evolution for $T_{\perp}$ is shown in dashed gray line, and the ultrarelativistic evolution for $T_{\perp}$ (cf. eqn. \eqref{eq:Pperp_compressing_UR}) is shown in solid black line. In panels $b$ and $d$, the nonrelativistic CGL evolution for $T_{\parallel}$ is shown in dashed gray line, and the ultrarelativistic evolution for $T_{\parallel}$ (cf. eqn. \eqref{eq:Ppar_compressing_UR}) is shown in solid black line.}
    \label{fig:beta_mime_comparison_compressing}
\end{figure}

For our compressing simulations, we also tested the dependency of our results with mass ratio and initial plasma beta. We tried the same values as in the shearing case, namely, $m_i/m_e = 8, 64, $ and $1836$ and $\beta_i^{\text{init}} = 0.05, 0.5$, and $5$. All simulations have initial $k_BT/m_ic^2=30$, and we are only showing ion results, as electron show a analogous behavior. The results are shown in figure \ref{fig:beta_mime_comparison_compressing}.

We can see that for the three mass ratio considered, the evolutions of the perpendicular and parallel pressures (panels $a$ and $b$ in fig. \ref{fig:beta_mime_comparison_compressing}) are well described by the ultrarelativistic equations \eqref{eq:Pperp_compressing_UR} and \eqref{eq:Ppar_compressing_UR}, respectively. We observe a similar pattern for the simulations with different initial beta (panels $c$ and $d$ in fig. \ref{fig:beta_mime_comparison_compressing}). In this case, both $T_{\perp,i}$ and $T_{\parallel,i}$ evolutions are well explained by equations \eqref{eq:Pperp_compressing_UR} and \eqref{eq:Ppar_compressing_UR} for all the values of beta considered. Therefore, similar to the nonrelativistic case, the relativistic extension of the double-adiabatic equation shows a robust independence of both mass ratio and initial plasma beta. This independence of plasma beta allows the application of these extensions to the double-adiabatic equations to a wide variety of astrophysical scenarios in which one or more species are relativistically hot.

\bibliographystyle{jpp}
\bibliography{jpp-instructions}

@INPROCEEDINGS{Kulsrud1983,
       author = {{Kulsrud}, R.~M.},
        title = "{MHD description of plasma}",
    booktitle = {Basic Plasma Physics: Selected Chapters, Handbook of Plasma Physics, Volume 1},
         year = 1983,
       editor = {{Galeev}, A.~A. and {Sudan}, R.~N.},
        month = jan,
        pages = {1},
       adsurl = {https://ui.adsabs.harvard.edu/abs/1983bpp..conf....1K}
}

@BOOK{Kulsrud2005,
       author = {{Kulsrud}, Russell M.},
        title = "{Plasma Physics for Astrophysics}",
         year = 2005,
      address = {Princeton},
    publisher = {Princeton Univ. Press},
}

@article{CGL1956,
 ISSN = {00804630},
 URL = {http://www.jstor.org/stable/99870},
 author = {G. F. Chew and M. L. Goldberger and F. E. Low},
 journal = {Proceedings of the Royal Society of London. Series A, Mathematical and Physical Sciences},
 number = {1204},
 pages = {112--118},
 publisher = {Royal Society},
 title = {The Boltzmann Equation and the One-Fluid Hydromagnetic Equations in the Absence of Particle Collisions},
 urldate = {2026-01-13},
 volume = {236},
 year = {1956}
}

@BOOK{RezzollaZanotti2013,
       author = {{Rezzolla}, Luciano and {Zanotti}, Olindo},
        title = "{Relativistic Hydrodynamics}",
         year = 2013,
       adsurl = {https://ui.adsabs.harvard.edu/abs/2013rehy.book.....R},
      publisher = {Oxford University Press}
}

@inbook{Buneman1993,
  title={Computer Space Plasma Physics},
  chapter={3},
  pages={67},
  author={Buneman, O.},
  isbn={9784887041110},
  year={1993},
  publisher={Terra Scientific Publishing Company, Tokyo}
}

@article{Spitkovsky2005,
author = {Spitkovsky,Anatoly },
title = {Simulations of relativistic collisionless shocks: shock structure and particle acceleration},
journal = {AIP Conference Proceedings},
volume = {801},
number = {1},
pages = {345-350},
year = {2005},
}

@article{Riquelme2012,
	year = 2012,
	month = {jul},
	publisher = {American Astronomical Society},
	volume = {755},
	number = {1},
	pages = {50},
	author = {Mario A. Riquelme and Eliot Quataert and Prateek Sharma and Anatoly Spitkovsky},
	title = {{LOCAL} {TWO}-{DIMENSIONAL} {PARTICLE}-{IN}-{CELL} {SIMULATIONS} {OF} {THE} {COLLISIONLESS} {MAGNETOROTATIONAL} {INSTABILITY}},
	journal = {The Astrophysical Journal},
}

@article{Tran2023,
doi = {10.3847/1538-4357/acbef9},
url = {https://doi.org/10.3847/1538-4357/acbef9},
year = {2023},
month = {may},
publisher = {The American Astronomical Society},
volume = {948},
number = {2},
pages = {130},
author = {Tran, Aaron and Sironi, Lorenzo and Ley, Francisco and Zweibel, Ellen G. and Riquelme, Mario A.},
title = {Electron Reacceleration via Ion Cyclotron Waves in the Intracluster Medium},
journal = {The Astrophysical Journal},
}

@article{Barnes1966,
    author = {Barnes, Aaron},
    title = {Collisionless Damping of Hydromagnetic Waves},
    journal = {The Physics of Fluids},
    volume = {9},
    number = {8},
    pages = {1483-1495},
    year = {1966},
    month = {08},
    issn = {0031-9171},
    doi = {10.1063/1.1761882},
    url = {https://doi.org/10.1063/1.1761882},
}

@article{Hasegawa1969,
    author = {Hasegawa, Akira},
    title = {Drift Mirror Instability in the Magnetosphere},
    journal = {The Physics of Fluids},
    volume = {12},
    number = {12},
    pages = {2642-2650},
    year = {1969},
    month = {12},
    issn = {0031-9171},
    doi = {10.1063/1.1692407},
    url = {https://doi.org/10.1063/1.1692407},
}

@article{vedenov1959,
  title={Plasma Physics and Problem of Controlled Thermonuclear Reactions, Vol. III, ed. MA Leontovich, 332-339},
  author={Vedenov, AA and Sagdeev, RZ},
  journal={English edition, Pergamon, NY},
  year={1959}
}

@article{rudakov1959,
  title={Plasma Physics and the Problem of Controlled Thermonuclear Reactions},
  author={Rudakov, LI and Sagdeev, RZ},
  journal={V},
  volume={3},
  pages={321},
  year={1959}
}

@article{Chandra1958,
    author = {Chandrasekhar, Subrahmanyan and Kaufman, A. N. and Watson, K. M.},
    title = {The stability of the pinch},
    journal = {Proceedings of the Royal Society of London. A. Mathematical and Physical Sciences},
    volume = {245},
    number = {1243},
    pages = {435-455},
    year = {1958},
    month = {07},
    issn = {0080-4630},
    doi = {10.1098/rspa.1958.0094},
    url = {https://doi.org/10.1098/rspa.1958.0094},
}

@article{SouthwoodKivelson1993,
author = {Southwood, David J. and Kivelson, Margaret G.},
title = {Mirror instability: 1. Physical mechanism of linear instability},
journal = {Journal of Geophysical Research: Space Physics},
volume = {98},
number = {A6},
pages = {9181-9187},
doi = {https://doi.org/10.1029/92JA02837},
url = {https://agupubs.onlinelibrary.wiley.com/doi/abs/10.1029/92JA02837},
year = {1993}
}

@article{KivelsonSouthwood1996,
author = {Kivelson, Margaret Galland and Southwood, David J.},
title = {Mirror instability II: The mechanism of nonlinear saturation},
journal = {Journal of Geophysical Research: Space Physics},
volume = {101},
number = {A8},
pages = {17365-17371},
doi = {https://doi.org/10.1029/96JA01407},
year = {1996}
}

@article{Pokhotelov2004,
author = {Pokhotelov, O. A. and Sagdeev, R. Z. and Balikhin, M. A. and Treumann, R. A.},
title = {Mirror instability at finite ion-Larmor radius wavelengths},
journal = {Journal of Geophysical Research: Space Physics},
volume = {109},
number = {A9},
pages = {},
doi = {https://doi.org/10.1029/2004JA010568},
year = {2004}
}

@article{Pokhotelov2002,
author = {Pokhotelov, Oleg A. and Treumann, Rudolf A. and Sagdeev, Roald Z. and Balikhin, Michael A. and Onishchenko, Oleg G. and Pavlenko, Vladimir P. and Sandberg, Ingmar},
title = {Linear theory of the mirror instability in non-Maxwellian space plasmas},
journal = {Journal of Geophysical Research: Space Physics},
volume = {107},
number = {A10},
pages = {SMP 18-1-SMP 18-11},
doi = {https://doi.org/10.1029/2001JA009125},
year = {2002}
}

@article{Parker1958,
  title = {Dynamical Instability in an Anisotropic Ionized Gas of Low Density},
  author = {Parker, E. N.},
  journal = {Phys. Rev.},
  volume = {109},
  issue = {6},
  pages = {1874--1876},
  numpages = {0},
  year = {1958},
  month = {Mar},
  publisher = {American Physical Society},
  doi = {10.1103/PhysRev.109.1874},
  url = {https://link.aps.org/doi/10.1103/PhysRev.109.1874}
}

@INPROCEEDINGS{VedenovSagdeev1961,
       author = {{Vedenov}, A.~A. and {Sagdeev}, R.~Z.},
        title = "{Some properties of a plasma with an anisotropic ion velocity distribution in a magnetic field}",
    booktitle = {Plasma Physics and the Problem of Controlled Thermonuclear Reactions, Volume 3},
         year = 1961,
       editor = {{Leontovich}, M.~A.},
       volume = {3},
        month = jan,
        pages = {332},
       adsurl = {https://ui.adsabs.harvard.edu/abs/1961ppp3.conf..332V}
}

@article{Gary1998,
author = {Gary, S. Peter and Li, Hui and O'Rourke, Sean and Winske, Dan},
title = {Proton resonant firehose instability: Temperature anisotropy and fluctuating field constraints},
journal = {Journal of Geophysical Research: Space Physics},
volume = {103},
number = {A7},
pages = {14567-14574},
doi = {https://doi.org/10.1029/98JA01174},
url = {https://agupubs.onlinelibrary.wiley.com/doi/abs/10.1029/98JA01174},
year = {1998}
}

@article{Yoon1993,
    author = {Yoon, Peter H. and Wu, C. S. and de Assis, A. S.},
    title = {Effect of finite ion gyroradius on the fire‐hose instability in a high beta plasma},
    journal = {Physics of Fluids B: Plasma Physics},
    volume = {5},
    number = {7},
    pages = {1971-1979},
    year = {1993},
    month = {07},
    issn = {0899-8221},
    doi = {10.1063/1.860785},
    url = {https://doi.org/10.1063/1.860785},
}

@article{HellingerMatsumoto2000,
author = {Hellinger, P. and Matsumoto, H.},
title = {New kinetic instability: Oblique Alfvén fire hose},
journal = {Journal of Geophysical Research: Space Physics},
volume = {105},
number = {A5},
pages = {10519-10526},
doi = {https://doi.org/10.1029/1999JA000297},
year = {2000}
}

@article{LiHabbal2000,
author = {Li, Xing and Habbal, Shadia Rifai},
title = {Electron kinetic firehose instability},
journal = {Journal of Geophysical Research: Space Physics},
volume = {105},
number = {A12},
pages = {27377-27385},
doi = {https://doi.org/10.1029/2000JA000063},
year = {2000}
}

@article{GaryNishimura2003,
    author = {Gary, S. Peter and Nishimura, Kazumi},
    title = {Resonant electron firehose instability: Particle-in-cell simulations},
    journal = {Physics of Plasmas},
    volume = {10},
    number = {9},
    pages = {3571-3576},
    year = {2003},
    month = {09},
    issn = {1070-664X},
    doi = {10.1063/1.1590982},
    url = {https://doi.org/10.1063/1.1590982},
}

@article{SagdeevShafranov1960,
author = {Sagdeev, R.Z. and Shafranov, V.D.},
title = {On the stability of a plasma with an anisotropic distribution of velocities in a magnetic field},
journal = {J. Exptl. Theoret. Phys.},
volume = {39},
pages = {181-184},
year = {1960}
}

@article{Gary1992,
author = {Gary, S. Peter},
title = {The mirror and ion cyclotron anisotropy instabilities},
journal = {Journal of Geophysical Research: Space Physics},
volume = {97},
number = {A6},
pages = {8519-8529},
doi = {https://doi.org/10.1029/92JA00299},
url = {https://agupubs.onlinelibrary.wiley.com/doi/abs/10.1029/92JA00299},
year = {1992}
}

@article{Gary1993,
author = {Gary, S. Peter and Fuselier, Stephen A. and Anderson, Brian J.},
title = {Ion anisotropy instabilities in the magnetosheath},
journal = {Journal of Geophysical Research: Space Physics},
volume = {98},
number = {A2},
pages = {1481-1488},
doi = {https://doi.org/10.1029/92JA01844},
year = {1993}
}

@article{Lopez2016,
doi = {10.3847/0004-637X/832/1/36},
url = {https://doi.org/10.3847/0004-637X/832/1/36},
year = {2016},
month = {nov},
publisher = {The American Astronomical Society},
volume = {832},
number = {1},
pages = {36},
author = {López, Rodrigo A. and Moya, Pablo S. and Navarro, Roberto E. and Araneda, Jaime A. and Muñoz, Víctor and Viñas, Adolfo F. and Valdivia, J. Alejandro},
title = {RELATIVISTIC CYCLOTRON INSTABILITY IN ANISOTROPIC PLASMAS},
journal = {The Astrophysical Journal},
}

@article{HellingerStverak2018, title={Electron mirror instability: particle-in-cell simulations}, volume={84}, DOI={10.1017/S0022377818000703}, number={4}, journal={Journal of Plasma Physics}, author={Hellinger, Petr and Štverák, Štěpán}, year={2018}, pages={905840402}}

@article{Lopez2022,
doi = {10.3847/1538-4357/ac66e4},
url = {https://doi.org/10.3847/1538-4357/ac66e4},
year = {2022},
month = {may},
publisher = {The American Astronomical Society},
volume = {930},
number = {2},
pages = {158},
author = {López, R. A. and Micera, A. and Lazar, M. and Poedts, S. and Lapenta, G. and Zhukov, A. N. and Boella, E. and Shaaban, S. M.},
title = {Mixing the Solar Wind Proton and Electron Scales. Theory and 2D-PIC Simulations of Firehose Instability},
journal = {The Astrophysical Journal},
}

@article{Gary1996,
author = {Gary, S. Peter and Wang, Joseph},
title = {Whistler instability: Electron anisotropy upper bound},
journal = {Journal of Geophysical Research: Space Physics},
volume = {101},
number = {A5},
pages = {10749-10754},
doi = {https://doi.org/10.1029/96JA00323},
url = {https://agupubs.onlinelibrary.wiley.com/doi/abs/10.1029/96JA00323},
year = {1996}
}

@article{KennelPetschek1966,
author = {Kennel, C. F. and Petschek, H. E.},
title = {Limit on stably trapped particle fluxes},
journal = {Journal of Geophysical Research (1896-1977)},
volume = {71},
number = {1},
pages = {1-28},
doi = {https://doi.org/10.1029/JZ071i001p00001},
year = {1966}
}

@article{Bott2024, title={Kinetic stability of Chapman–Enskog plasmas}, volume={90}, DOI={10.1017/S0022377824000308}, number={2}, journal={Journal of Plasma Physics}, author={Bott, Archie F.A. and Cowley, S.C. and Schekochihin, A.A.}, year={2024}, pages={975900207}}

@article{Kunz2014,
  title = {Firehose and Mirror Instabilities in a Collisionless Shearing Plasma},
  author = {Kunz, Matthew W. and Schekochihin, Alexander A. and Stone, James M.},
  journal = {Phys. Rev. Lett.},
  volume = {112},
  issue = {20},
  pages = {205003},
  numpages = {6},
  year = {2014},
  month = {May},
  publisher = {American Physical Society},
  doi = {10.1103/PhysRevLett.112.205003},
  url = {https://link.aps.org/doi/10.1103/PhysRevLett.112.205003}
}

@article{Riquelme2015,
doi = {10.1088/0004-637X/800/1/27},
url = {https://doi.org/10.1088/0004-637X/800/1/27},
year = {2015},
month = {feb},
publisher = {The American Astronomical Society},
volume = {800},
number = {1},
pages = {27},
author = {Riquelme, Mario A. and Quataert, Eliot and Verscharen, Daniel},
title = {PARTICLE-IN-CELL SIMULATIONS OF CONTINUOUSLY DRIVEN MIRROR AND ION CYCLOTRON INSTABILITIES IN HIGH BETA ASTROPHYSICAL AND HELIOSPHERIC PLASMAS},
journal = {The Astrophysical Journal},
}

@article{Riquelme2016,
doi = {10.3847/0004-637X/824/2/123},
url = {https://doi.org/10.3847/0004-637X/824/2/123},
year = {2016},
month = {jun},
publisher = {The American Astronomical Society},
volume = {824},
number = {2},
pages = {123},
author = {Riquelme, Mario A. and Quataert, Eliot and Verscharen, Daniel},
title = {{PIC} SIMULATIONS OF THE EFFECT OF VELOCITY SPACE INSTABILITIES ON ELECTRON VISCOSITY AND THERMAL CONDUCTION},
journal = {The Astrophysical Journal},
}

@article{Riquelme2018,
doi = {10.3847/1538-4357/aaa6d1},
url = {https://doi.org/10.3847/1538-4357/aaa6d1},
year = {2018},
month = {feb},
publisher = {The American Astronomical Society},
volume = {854},
number = {2},
pages = {132},
author = {Riquelme, Mario and Quataert, Eliot and Verscharen, Daniel},
title = {PIC Simulations of Velocity-space Instabilities in a Decreasing Magnetic Field: Viscosity and Thermal Conduction},
journal = {The Astrophysical Journal},
}

@article{Innocenti2019,
doi = {10.3847/1538-4357/ab3e40},
url = {https://doi.org/10.3847/1538-4357/ab3e40},
year = {2019},
month = {sep},
publisher = {The American Astronomical Society},
volume = {883},
number = {2},
pages = {146},
author = {Innocenti, Maria Elena and Tenerani, Anna and Boella, Elisabetta and Velli, Marco},
title = {Onset and Evolution of the Oblique, Resonant Electron Firehose Instability in the Expanding Solar Wind Plasma},
journal = {The Astrophysical Journal},
}

@article{Ley2024,
doi = {10.3847/1538-4357/ad2455},
url = {https://doi.org/10.3847/1538-4357/ad2455},
year = {2024},
month = {apr},
publisher = {The American Astronomical Society},
volume = {965},
number = {2},
pages = {155},
author = {Ley, Francisco and Zweibel, Ellen G. and Miller, Drake and Riquelme, Mario},
title = {Secondary Whistler and Ion-cyclotron Instabilities Driven by Mirror Modes in Galaxy Clusters},
journal = {The Astrophysical Journal},
}

@article{Bott2021,
doi = {10.3847/2041-8213/ac37c2},
url = {https://doi.org/10.3847/2041-8213/ac37c2},
year = {2021},
month = {nov},
publisher = {The American Astronomical Society},
volume = {922},
number = {2},
pages = {L35},
author = {Bott, A. F. A. and Arzamasskiy, L. and Kunz, M. W. and Quataert, E. and Squire, J.},
title = {Adaptive Critical Balance and Firehose Instability in an Expanding, Turbulent, Collisionless Plasma},
journal = {The Astrophysical Journal Letters},
}

@article{Zhdankin2023,
doi = {10.3847/1538-4357/acaf54},
url = {https://doi.org/10.3847/1538-4357/acaf54},
year = {2023},
month = {feb},
publisher = {The American Astronomical Society},
volume = {944},
number = {1},
pages = {24},
author = {Zhdankin, Vladimir and Kunz, Matthew W. and Uzdensky, Dmitri A.},
title = {Synchrotron Firehose Instability},
journal = {The Astrophysical Journal},
}

@article{Gary2011,
    author = {Gary, S. Peter and Liu, Kaijun and Winske, Dan},
    title = {Whistler anisotropy instability at low electron $\beta$: Particle-in-cell simulations},
    journal = {Physics of Plasmas},
    volume = {18},
    number = {8},
    pages = {082902},
    year = {2011},
    month = {08},
}

@article{Sironi2015,
doi = {10.1088/0004-637X/800/2/88},
url = {https://doi.org/10.1088/0004-637X/800/2/88},
year = {2015},
month = {feb},
publisher = {The American Astronomical Society},
volume = {800},
number = {2},
pages = {88},
author = {Sironi, Lorenzo and Narayan, Ramesh},
title = {ELECTRON HEATING BY THE ION CYCLOTRON INSTABILITY IN COLLISIONLESS ACCRETION FLOWS. I. COMPRESSION-DRIVEN INSTABILITIES AND THE ELECTRON HEATING MECHANISM},
journal = {The Astrophysical Journal},
}

@article{Zweibel2020,
doi = {10.3847/1538-4357/ab67bf},
url = {https://doi.org/10.3847/1538-4357/ab67bf},
year = {2020},
month = {feb},
publisher = {The American Astronomical Society},
volume = {890},
number = {1},
pages = {67},
author = {Zweibel, Ellen G.},
title = {The Role of Pressure Anisotropy in Cosmic-Ray Hydrodynamics},
journal = {The Astrophysical Journal},
}

@article{Akiyama2019,
doi = {10.3847/2041-8213/ab0f43},
url = {https://doi.org/10.3847/2041-8213/ab0f43},
year = {2019},
month = {apr},
publisher = {The American Astronomical Society},
volume = {875},
number = {1},
pages = {L5},
author = {{EHT Collaboration}}, 
title = {First {M87} {E}vent {H}orizon {T}elescope {R}esults. {V}. {P}hysical {O}rigin of the {A}symmetric {R}ing},
journal = {The Astrophysical Journal Letters},
}

@article{Verscharen2018, title={ALPS: the Arbitrary Linear Plasma Solver}, volume={84}, DOI={10.1017/S0022377818000739}, number={4}, journal={Journal of Plasma Physics}, author={Verscharen, D. and Klein, K.G. and Chandran, B.D.G. and Stevens, M.L. and Salem, C.S. and Bale, S.D.}, year={2018}, pages={905840403}}

@article{Majeski2024, title={Self-organization in collisionless, high-$\beta$ turbulence}, volume={90}, DOI={10.1017/S0022377824001296}, number={6}, journal={Journal of Plasma Physics}, author={Majeski, S. and Kunz, M.W. and Squire, J.}, year={2024}, pages={535900601}}

@article{Hunana2019, title={An introductory guide to fluid models with anisotropic temperatures. Part 1. {CGL} description and collisionless fluid hierarchy}, volume={85}, DOI={10.1017/S0022377819000801}, number={6}, journal={Journal of Plasma Physics}, author={Hunana, P. and Tenerani, A. and Zank, G. P. and Khomenko, E. and Goldstein, M. L. and Webb, G. M. and Cally, P. S. and Collados, M. and Velli, M. and Adhikari, L.}, year={2019}, pages={205850602}}

@article{Gedalin1991,
    author = {Gedalin, M.},
    title = {Relativistic hydrodynamics and thermodynamics of anisotropic plasmas},
    journal = {Physics of Fluids B: Plasma Physics},
    volume = {3},
    number = {8},
    pages = {1871-1875},
    year = {1991},
    month = {08},
    issn = {0899-8221},
    doi = {10.1063/1.859656},
    url = {https://doi.org/10.1063/1.859656},
}

@article{GedalinOiberman,
  title = {Generally covariant relativistic anisotropic magnetohydrodynamics},
  author = {Gedalin, M. and Oiberman, I.},
  journal = {Phys. Rev. E},
  volume = {51},
  issue = {5},
  pages = {4901--4907},
  numpages = {0},
  year = {1995},
  month = {May},
  publisher = {American Physical Society},
  doi = {10.1103/PhysRevE.51.4901},
  url = {https://link.aps.org/doi/10.1103/PhysRevE.51.4901}
}

@article{Matteini2013,
author = {Matteini, Lorenzo and Hellinger, Petr and Goldstein, Bruce E. and Landi, Simone and Velli, Marco and Neugebauer, Marcia},
title = {Signatures of kinetic instabilities in the solar wind},
journal = {Journal of Geophysical Research: Space Physics},
volume = {118},
number = {6},
pages = {2771-2782},
doi = {https://doi.org/10.1002/jgra.50320},
url = {https://agupubs.onlinelibrary.wiley.com/doi/abs/10.1002/jgra.50320},
year = {2013}
}

@article{Matteini2007,
author = {Matteini, Lorenzo and Landi, Simone and Hellinger, Petr and Pantellini, Filippo and Maksimovic, Milan and Velli, Marco and Goldstein, Bruce E. and Marsch, Eckart},
title = {Evolution of the solar wind proton temperature anisotropy from 0.3 to 2.5 AU},
journal = {Geophysical Research Letters},
volume = {34},
number = {20},
pages = {},
doi = {https://doi.org/10.1029/2007GL030920},
url = {https://agupubs.onlinelibrary.wiley.com/doi/abs/10.1029/2007GL030920},
year = {2007}
}

@article{Chandra2015,
doi = {10.1088/0004-637X/810/2/162},
url = {https://doi.org/10.1088/0004-637X/810/2/162},
year = {2015},
month = {sep},
publisher = {The American Astronomical Society},
volume = {810},
number = {2},
pages = {162},
author = {Chandra, Mani and Gammie, Charles F. and Foucart, Francois and Quataert, Eliot},
title = {AN EXTENDED MAGNETOHYDRODYNAMICS MODEL FOR RELATIVISTIC WEAKLY COLLISIONAL PLASMAS},
journal = {The Astrophysical Journal},
}

@article{TenBarge2008,
    author = {TenBarge, J. M. and Hazeltine, R. D. and Mahajan, S. M.},
    title = {Fluid model for relativistic, magnetized plasmas},
    journal = {Physics of Plasmas},
    volume = {15},
    number = {6},
    pages = {062112},
    year = {2008},
    month = {06},
    issn = {1070-664X},
    doi = {10.1063/1.2937123},
}

@ARTICLE{Braginskii1965,
       author = {{Braginskii}, S.~I.},
        title = "{Transport Processes in a Plasma}",
      journal = {Reviews of Plasma Physics},
         year = 1965,
        month = jan,
       volume = {1},
        pages = {205},
}

@article{Newcomb1982,
    author = {Newcomb, William A.},
    title = {Warm relativistic electron fluid},
    journal = {The Physics of Fluids},
    volume = {25},
    number = {5},
    pages = {846-851},
    year = {1982},
    month = {05},
    issn = {0031-9171},
    doi = {10.1063/1.863814},
    }

@article{HolmKuperschmidt1986,
    author = {Holm, Darryl D. and Kupershmidt, Boris A.},
    title = {Hamiltonian theory of relativistic magnetohydrodynamics with anisotropic pressure},
    journal = {The Physics of Fluids},
    volume = {29},
    number = {11},
    pages = {3889-3891},
    year = {1986},
    month = {11},
    issn = {0031-9171},
    doi = {10.1063/1.865774},
}

@ARTICLE{Wierzchucka2026,
       author = {{Wierzchucka}, Agnieszka and {Bilbao}, Pablo J. and {Thomas}, Alexander G.~R. and {Uzdensky}, Dmitri A. and {Schekochihin}, Alexander A.},
        title = "{Double-Adiabatic Equations of State for Relativistic Plasmas}",
      journal = {arXiv e-prints},
         year = 2026,
        month = mar,
          eid = {arXiv:2603.25669},
        pages = {arXiv:2603.25669},
          doi = {10.48550/arXiv.2603.25669},
archivePrefix = {arXiv},
       eprint = {2603.25669},
 primaryClass = {astro-ph.HE},
       adsurl = {https://ui.adsabs.harvard.edu/abs/2026arXiv260325669W}
}

@article{Lichko2017,
doi = {10.3847/2041-8213/aa9a33},
url = {https://doi.org/10.3847/2041-8213/aa9a33},
year = {2017},
month = {nov},
publisher = {The American Astronomical Society},
volume = {850},
number = {2},
pages = {L28},
author = {Lichko, E. and Egedal, J. and Daughton, W. and Kasper, J.},
title = {Magnetic Pumping as a Source of Particle Heating and Power-law Distributions in the Solar Wind},
journal = {The Astrophysical Journal Letters},
}

\end{document}